\documentclass[twocolumn]{aastex631}

\shorttitle{\texttt{democratic detrender}}
\shortauthors{Yahalomi et al.}
\graphicspath{{./}{figures/}}
\begin{document}

\title{The \texttt{democratic detrender}: \linebreak Ensemble-Based Removal of the Nuisance Signal in Stellar Time-Series Photometry}

\correspondingauthor{Daniel A. Yahalomi}
\email{dyahalomi@flatironinstitute.org}

\author[0000-0003-4755-584X]{Daniel A. Yahalomi}\thanks{Flatiron Research Fellow}
\affiliation{Department of Astronomy, Columbia University, 550 W 120th St., New York NY 10027, USA}
\affiliation{Center for Computational Astrophysics, Flatiron Institute, 162 Fifth Ave, New York, NY 10010, USA}

\author[0000-0002-4365-7366]{David Kipping} 
\affiliation{Department of Astronomy, Columbia University, 550 W 120th St., New York NY 10027, USA}

\author[0000-0003-0360-4451]{Diana Solano-Oropeza} 
\affiliation{Department of Astronomy, Columbia University, 550 W 120th St., New York NY 10027, USA}
\affiliation{Department of Astronomy and Carl Sagan Institute, Cornell University, 122 Sciences Drive, Ithaca, NY 14850, USA}

\author[0000-0003-2866-1436]{Madison Li} 
\affiliation{Department of Astronomy, Barnard College, 3009 Broadway, New York NY 10027, USA}
\affiliation{Department of Astronomy, Columbia University, 550 W 120th St., New York NY 10027, USA}

\author[0009-0000-5314-5770]{Avishi Poddar} 
\affiliation{Department of Astronomy, Columbia University, 550 W 120th St., New York NY 10027, USA}

\author[0009-0008-5025-9818]{Xunhe (Andrew) Zhang} 
\affiliation{Department of Astronomy, Columbia University, 550 W 120th St., New York NY 10027, USA}

\author{Yassine Abaakil} 
\affiliation{Department of Astronomy, Columbia University, 550 W 120th St., New York NY 10027, USA}

\author[0000-0002-9544-0118]{Ben Cassese} 
\affiliation{Department of Astronomy, Columbia University, 550 W 120th St., New York NY 10027, USA}

\author[0000-0002-7032-2350]{Jeff Jennings}
\affiliation{Center for Computational Astrophysics, Flatiron Institute, 162 Fifth Ave, New York, NY 10010, USA}

\author[0009-0009-6718-6595]{Skylar Larsen}
\affiliation{Department of Astronomy and Carl Sagan Institute, Cornell University, 122 Sciences Drive, Ithaca, NY 14850, USA}

\author[0000-0001-7836-1787]{Jake D. Turner}
\affiliation{Department of Astronomy and Carl Sagan Institute, Cornell University, 122 Sciences Drive, Ithaca, NY 14850, USA}

\author[0000-0003-2331-5606]{Alex Teachey}
\affiliation{Institute of Astronomy and Astrophysics, Academia Sinica, Taipei 10617, Taiwan}

\author{Jiajing Liu}\thanks{Harvard SRMP Student} 
\affiliation{Cambridge Rindge and Latin School, Cambridge, MA 02138, USA}

\author{Farai Sundai}\thanks{Harvard SRMP Student} 
\affiliation{Cambridge Rindge and Latin School, Cambridge, MA 02138, USA}

\author{Lila Valaskovic}\thanks{Harvard SRMP Student}
\affiliation{Cambridge Rindge and Latin School, Cambridge, MA 02138, USA}



\begin{abstract}

Accurate, precise, and computationally efficient removal of unwanted activity that exists as a combination of periodic, quasi-periodic, and non-periodic systematic trends in time-series photometric data is a critical step in exoplanet transit analysis. Throughout the years, many different modeling methods have been used for this process, often called ``detrending.'' However, there is no community-wide consensus regarding the favored approach. In order to mitigate model dependency, we present an ensemble-based approach to detrending via community-of-models and the \texttt{democratic detrender}: a modular and scalable open-source coding package that implements ensemble detrending. The \texttt{democratic detrender} allows users to select from a number of packaged detrending methods (including cosine filtering, Gaussian processes, and polynomial fits) or provide their own set of detrended light curves via methods of their choosing. It then combines the individually detrended light curves into a single method marginalized light curve. Additionally, the \texttt{democratic detrender} inflates each data point’s uncertainty based on the scatter between detrenders, thereby propagating model-selection uncertainty into the final light curve. This ensemble strategy does not guarantee improvement over the single best-performing detrending method, but it substantially reduces the risk of selecting a detrending solution that is poorly calibrated or overfit to noise.


\end{abstract}

\keywords{}


\section{Introduction} \label{sec:intro}

Space-based stellar time-series photometry shows a combination of periodic, quasi-periodic, and non-periodic variations caused by both physical and observational factors. As discussed in a series of papers studying the noise properties in \textit{Kepler} time-series photometric data \citep[e.g.,][]{Jenkins2010, Gilliland2010, Gilliland2011, Gilliland2015},
there are four predominant components that together make up the observed variability: (1) detector noise, (2) stellar activity, (3) spacecraft motion, telescope, \& instrumental induced variations, and (4) the eclipses or transits of interest.

\begin{figure*}
    \centering 
    \includegraphics[width=\textwidth]{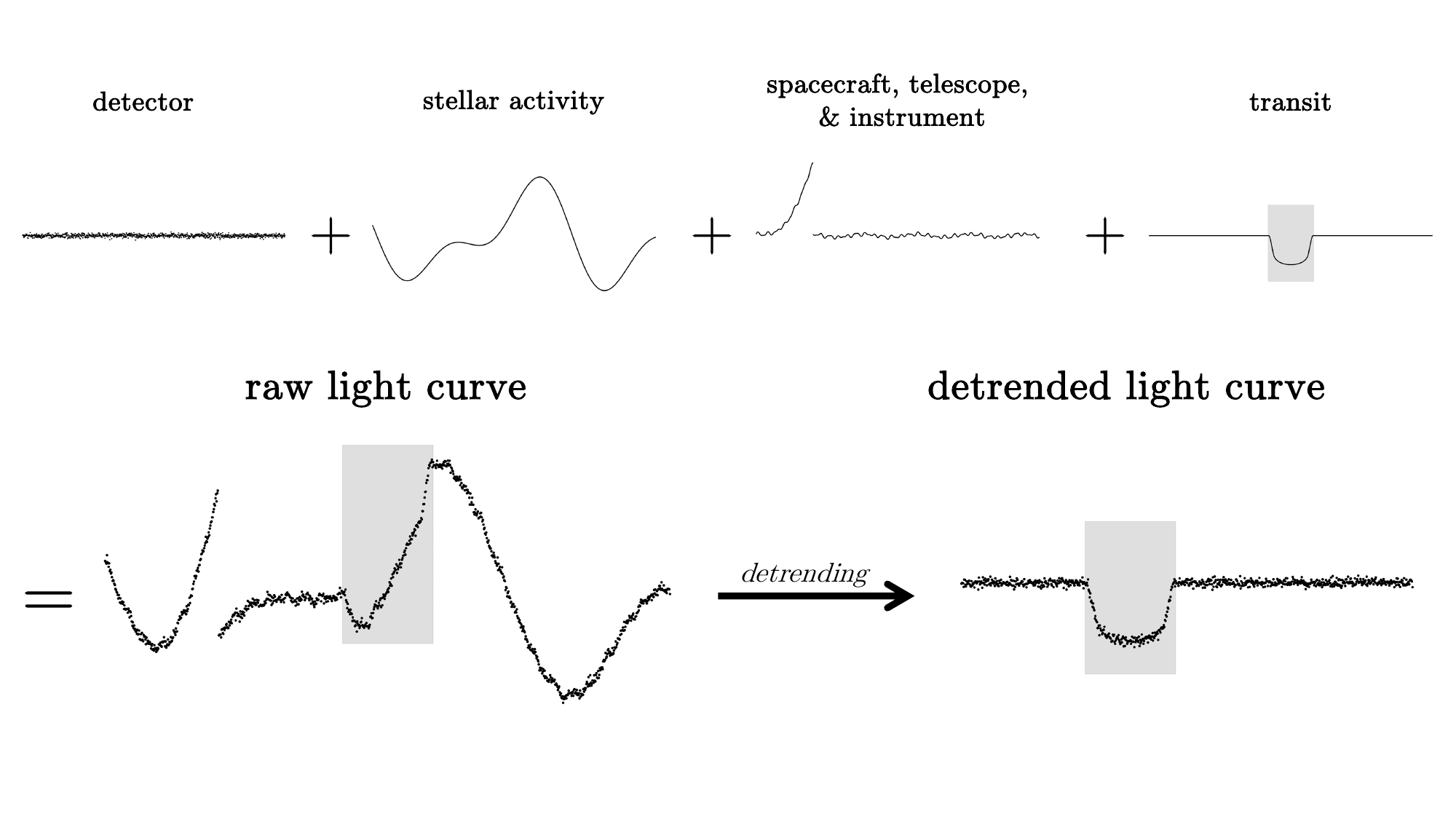}
    \caption{The four primary contributions to space-based stellar time-series photometric data are (1) detector noise, (2) stellar activity, (3) spacecraft motion, telescope, \& instrumental induced variations, and (4) eclipses or transits. Detrending is the process of removing as much of the stellar activity and spacecraft, telescope, \& instrument noise (i.e., ``nuisance signal'') as possible without adding additional non-physical features.}
    \label{fig: light curve}
\end{figure*}

Detector noise is non-periodic noise driven by random fluctuations in the detector and is typically assumed to be white noise. Stellar activity is a quasi-periodic phenomenon caused by fluctuations on the stellar surface. There are certain phenomena that on short timescales are largely periodic, such as star spots with a period equal to the rotational period of the star, and other phenomena, such as stellar flares, that are not expected to be periodic. Additionally, stellar photometry will always have some intrinsic Poisson noise that cannot be removed from the time-series light curves \citep{Schottky1918, Angus2017}. Spacecraft, telescope, \& instrument noise can be both period and non-periodic. Events such as momentum dumps for course correction happen on a periodic basis, but sudden changes caused by motion of the spacecraft, instrumental issues, or telescope issues do not follow a periodic schedule \citep[e.g.,][]{Borucki2010, Ricker2015}. Lastly, the eclipse or transit event will be close to periodic, as a single transit event occurs once per orbit of the planet. Stable planetary orbits will be purely periodic, albeit with minute fluctuations (or transit timing variations). An illustration of how these four different components contribute to space-based stellar time-series photometry can be seen in Figure~\ref{fig: light curve}.

As discussed in \citet{Hippke2019}, in the last decade, millions of light curves have been provided to the community thanks to space-based instruments such as \textit{CoRoT} \citep{Auvergne2009}, \textit{Kepler} \citep{Borucki2010}, \textit{K2} \citep{Howell2014}, \textit{TESS} \citep{Ricker2015}, and \textit{CHEOPS} \citep{Benz2021}. In the coming decade, this number will continue to grow, with future space-based time-series missions such as \textit{PLATO} \citep{Rauer2014}, \textit{Nancy Grace Roman Space Telescope} \citep{Spergel2015}, and \textit{Earth 2.0} \citep{Ge2022}. Ground based facilities, such as \textit{KELT} \citep{Pepper2007}, \textit{WASP} \citep{Pollacco2006}, and \textit{HATnet} \citep{Bakos2004}, have additionally contributed enormously to our wealth of time-series photometry and will continue to do so in the coming decade with instruments such as the \textit{Vera C. Rubin Observatory} \citep{Ivezic2019} coming online.

\begin{figure*}[htb!]
    \centering 
    \includegraphics[width=\textwidth]{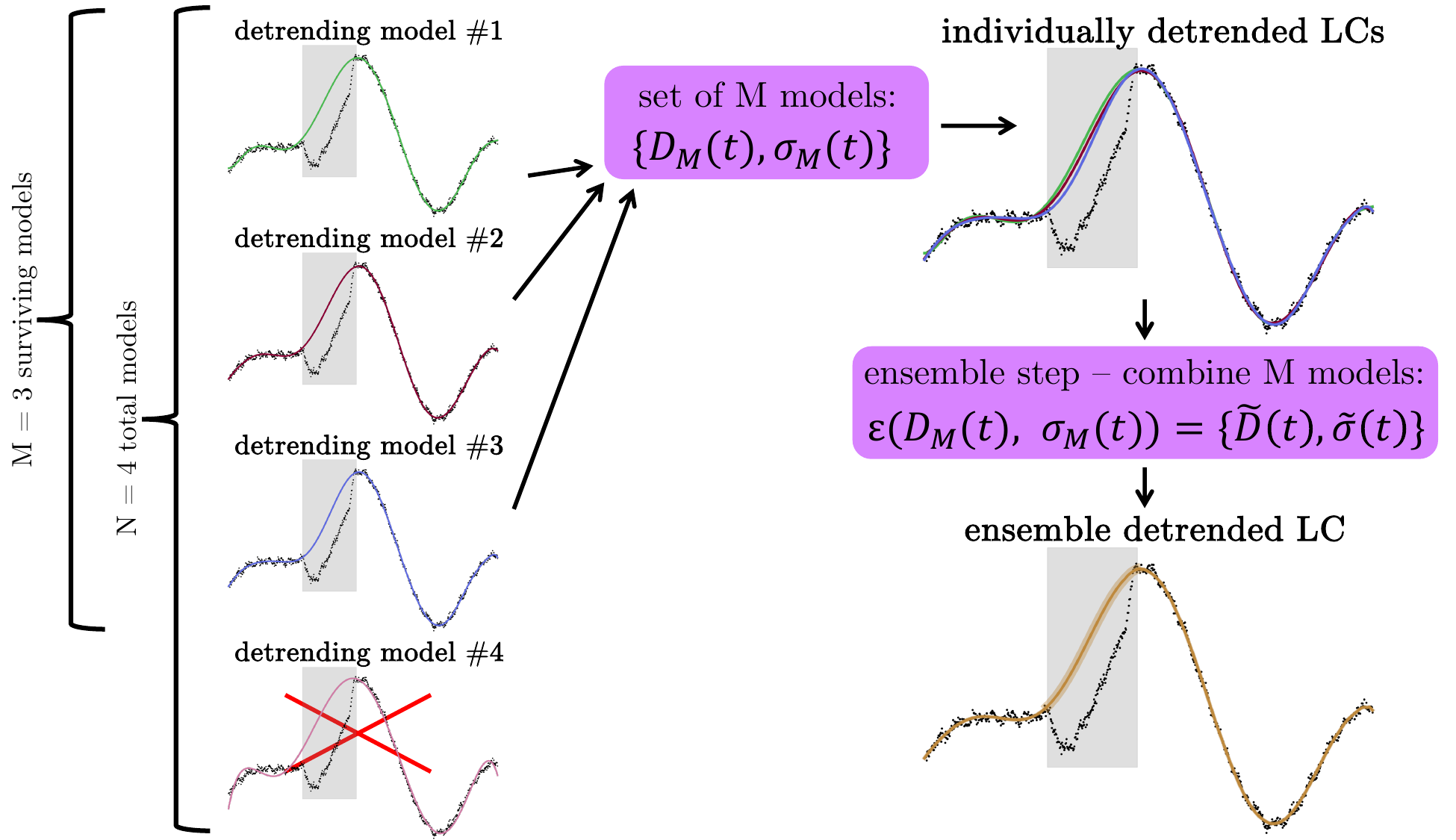}
    \caption{An example of ensemble detrending in practice. Here we've taken the simulated light curve (LC), presented in Figure~\ref{fig: light curve} and fit the non-transit data with four different detrending methods. Detrending model $\#$4 failed to accurately fit the data, and thus would fail the tests of Gaussianity and be removed before the ensemble step. In the top right one can see that detrending with each of the three individual detrended methods can result in noticeably different predictions for the transit shape. In the ensemble detrended light curve in the bottom right, we take the median detrended solution for each data point, as described in Equation~\ref{eq: ensemble median}. Additionally, we inflate the error bars, as shown by the shaded region in the ensemble detrended LC, by adding in quadrature the reported errors with the median absolute deviation between the different detrending methods, as described in Equation~\ref{eq: ensemble errors}. In total, the ensemble step is thus described by Equation~\ref{eq: ensemble function}.}
    \label{fig: ensemble detrending}
\end{figure*}

In modeling time-series photometry, particularly for short baseline eclipse events such as transiting exoplanets, eclipsing binaries, and self-lensing binaries, accurately and precisely removing coherent photometric variations is a critical step in correctly modeling the eclipse event. In response to the abundance of time-series data made available in the last decade many different modeling algorithms and techniques have been used for fitting out-of-transit trends in the light curves and removing these trends from the light curve -- this process will henceforth be called detrending (and is sometimes also referred to as pre-whitening) \citep[e.g.,][]{Mazeh2007, Kim2009, Mazeh2010, Ofir2010, Fabrycky2012, Gautier2012, Vanderburg2014, Waldmann2014, Aigrain2016, Luger2016, Angus2017, Sandford2017, Giles2018, Luger2018, Hippke2019, Morvan2020, Hattori2022, McGruder2022}. While many of these detrending methods have been found to work quite well in practice, from a theoretical standpoint they are by definition imperfect as they are all mispecified models. This is not a failure of the community to produce a ``perfect'' detrending models, but rather unfortunately the reality intrinsic to modeling stellar time-series photometry that is comprised of different periodic, quasi-periodic, and non-periodic components. This combined nuisance signal is so complicated that it is simply not solvable by any single detrending model. 

Precise and accurate detrending becomes even more critical when searching for transient events (e.g., exomoon transits) where one has to be especially careful to neither remove nor add the subtle signatures of the transient event that, unlike exoplanet transits, will not phase-fold coherently on the planetary period and thus one cannot expect morphological repetition \citep{TeacheyKipping2018}. A schematic example of detrending can also be seen in Figure~\ref{fig: light curve}.

The \texttt{democratic detrender} is written to be used with a method of transit fitting in which one models the transit in two steps: (1) modeling and removing unwanted activity and (2) fitting a transit to the detrending dataset. 

One might wonder about the alternative (but often computationally expensive) approach of jointly fitting, since our two-step approach admittedly has the disadvantage of assuming a perfect removal of unwanted activity in the subsequent transit modeling. In practice, this two-step method is only appropriate to use when the baseline of out-of-transit data greatly exceeds the baseline of in-transit data, a typical situation for transit surveys. This is because in the two-step method, one masks the in-transit data, trains (fits) the detrending models on the out-of-transit data, and then interpolates the detrending models over the in-transit times. As \textit{Kepler} and \textit{TESS} observe quarters or sectors for $\sim$month-long time periods, data from these missions are ideal for this method of detrending because they consistently have very long out-of-transit baselines. In contrast, if modeling a transit with a short baseline of data, such as typical \textit{JWST} or \textit{Hubble} observations, one would be better advised to model the nuisance signal in unison with the transit model. Additionally, when modeling \textit{K2} data, one would be better off using a detrending method explicitly designed to account for the large effects of the frequent thruster firings -- two of the most commonly used algorithms are self-flat-fielding (SFF) presented in \citet{Vanderburg2014} and the \texttt{EVEREST} package \citep{Luger2016, Luger2018}.

In three of the four detrending models (excluding the \texttt{local} model) used in the \texttt{democratic detrender} by default there is an emphasis on fitting low frequency (long period) noise and ignoring high frequency (short period) noise. This is so that we will not distort the transit shape with non-physical high frequency noise \citep{Waldmann2012}. Additionally, this decreases the chance of either hiding or erroneously adding moon-like transit dips -- i.e., minimizes both the false negative and false positive rates. For \textit{Kepler} and \textit{TESS} data, where we will frequently have multiple transit epochs, we would not expect high frequency noise to be coherent with the transits \citep{Pont2006}. Thus, by including multiple transits this incoherent high frequency noise should essentially cancel itself out, scaling with $\sqrt{N}$ transit epochs. In this context, the boundary between high frequency and low frequency noise is defined at a frequency commensurate with the transit duration.

In what follows, in Section~\ref{sec: ensemble detrending} we present the theory behind ensemble-based detrending via a community of models. Then, in Section~\ref{sec: detrending methods} we outline the detrending methods implemented by default in the \texttt{democratic detrender}, its modularity, and its customizability -- and walk through an example case of detrending a light curve with the \texttt{democratic detrender}. The source code can be found on GitHub.\footnote{\href{https://github.com/dyahalomi/democratic\_detrender}{https://github.com/dyahalomi/democratic\_detrender}} Tutorials and API documentation can be found on Read the Docs.\footnote{\url{https://democratic-detrender.readthedocs.io/en/latest/}} A static version of the code used in this work is available in Zenodo \citep{democratic_detrender_zenodo}.

\section{Ensemble Methods} \label{sec: ensemble detrending}

\subsection{Ensemble Learning}

As there exists no perfect detrending models, we adopt an approach similar to the common machine learning technique of ensemble learning, where a community of networks each classify a dataset and then the final result is a combination of the individual classifiers. Ensemble learning systems have become increasingly useful tools to improve the accuracy of automated decision-making systems \citep{ensemble_learning}. The underlying idea is that by aggregating the predictions of a group of models, one can often achieve better results than any single model could alone. As noted in \citet{ensemble_learning}, while ensemble-based decision making in machine learning is a relatively new concept of the past several decades \citep{DasarathySheela1979, HansenSalamon1990, Schapire1990}, ensemble-based decision making is a fundamental manner in which humans have made decisions for as long as civilized communities have existed. Mirroring the fundamental philosophy behind democracy, in which decision is made via a vote of the people, ensemble-based decision making is a process in which the final model values are decided via a ``vote'' of a set of included independent models. This strategy echoes the well-documented phenomenon observed by Francis Galton in 1907, who found that taking the median or the mean of many independent estimates—such as guesses of an ox’s weight at a county fair—could be remarkably accurate \citep{Galton1907}. This principle, now often cited as the ``wisdom of crowds,'' underpins the logic of ensemble decision-making: diversity and independence among estimators can lead to superior collective outcomes.

Ensemble methods take advantage of the fact that modeling error consists of two components: bias (systematic error, which affects accuracy) and variance (random error, which affects precision) \citep{ensemble_learning}. By combining multiple independent detrending models---each of which may suffer from different biases and overfitting tendencies---ensemble methods primarily improve accuracy by reducing the likelihood that a poorly performing or highly biased model dominates the final result. At the same time, averaging or median-selection across models can help mitigate variance, thereby reducing the risk of overfitting and improving the stability (precision) of the solution. Two key considerations arise in applying ensemble decision-making to detrending: (1) the choice of aggregation strategy (i.e., voting scheme) used to combine the outputs of different models, and (2) the recognition that while the ensemble may not always outperform the best individual model, it substantially lowers the risk of selecting a detrending solution that is poorly calibrated or overfit to noise \citep{ensemble_learning}.

\subsection{Ensemble Detrending}

\begin{figure*}
    \centering 
    \includegraphics[width=\textwidth]{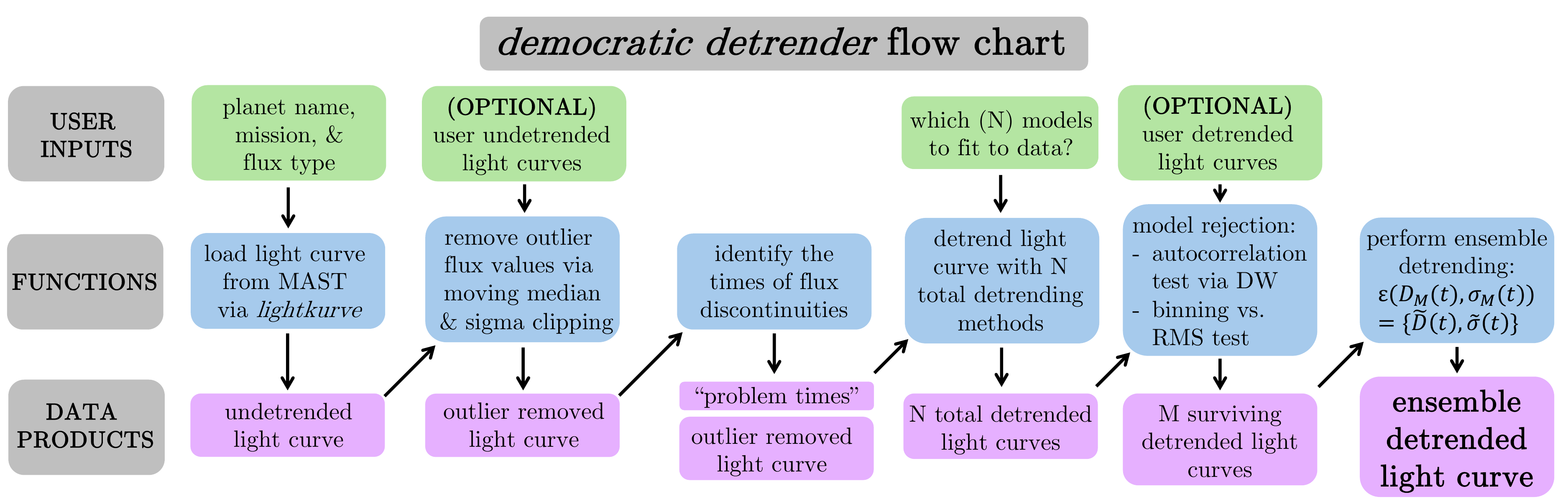}
    \caption{Flow chart depicting required and optional user inputs, built-in functions, and output data products for the \texttt{democratic detrender}. As long as the user passes in the requisite inputs and data products, the \texttt{democratic detrender} can be initiated at any of the intermediate functions. This flow chart thus displays the modularity and customizability of the \texttt{democratic detrender}.}
    \label{fig: flow chart}
\end{figure*}

In general, each of the individual detrending models are equally valued by the community, and so we have no reason to weight any models more than others. Additionally, as we want to exclude outliers from strongly influencing our detrended light curve values, we adopt a median selection for our ensemble voting step (although users can change this to mean selection). We additionally inflate the uncertainties in our final detrending light curve based on the variance between the different detrending methods -- allowing us to propagate information on the agreement between the different independent models. A schematic of ensemble detrending can be seen in Figure~\ref{fig: ensemble detrending}.

In order to not bias the detrending results, we emphasize that it is important to include detrending models that rely on different basis sets and different cost functions. For example, one wouldn't want to pass multiple polynomials of different orders into the ensemble detrender as these would all have the same basis sets and thus one would be essentially giving more weight to polynomial fits in the ensemble detrending light curve. The goal of ensemble detrending is to mitigate for model-dependent imperfections in the detrending, and so by using disparate detrending methods, one decreases the likelihood of biasing the final result. This principle echoes findings from social science research showing that groups composed of socially diverse individuals reach more accurate conclusions than homogeneous groups, even if they feel less confident in those decisions \citep{Phillips2009}. Similarly, in ensemble detrending, incorporating a diverse set of models — with different basis sets and cost functions — reduces the risk of systematic bias and leads to more robust and accurate results.

As stated previously, in its current form, the \texttt{democratic detrender} adopts an unweighted approach in combining the outputs of the included models. This unweighted assumption holds as long as the selected detrending models rely on different basis sets and cost functions and can each be trusted equally and independently. However, future versions could incorporate weighting schemes to account for varying levels of confidence or performance across the individual detrending models.

Assuming we start with N total models for the stellar activity per transit epoch, similar to \citet{Kipping2022}, we first test the Gaussianity of each of the N individually detrended light curves in two ways: (1) autocorrelation test at a timescale of the data's cadence via the Durbin-Watson statistic and (2) Poisson counting of independent measures test via binning vs. root-mean-square metric. These tests are described in much more detail in Section~\ref{sec: model rejection}. Once we remove anomalous ($>$ 3-$\sigma$ outliers; defined in Section~~\ref{sec: model rejection}) detrending models, we are then left with $M$ ($\leq$ $N$) surviving detrending models per epoch. We note that this 3-$\sigma$ cutoff removes the complete epoch of any detrending model that shows non-white noise properties -- rather than a point-by-point removal of $\sim$0.3\% of the data. Here, the priority is ensuring that we only pass white light curves into the final ensemble step, even if this means having fewer final detrended light curves passed into the ensemble step.

Let $D_m(t)$ denote the light curve at time $t$ produced by the $m$-th detrending model, where $m = 1, \dots, M$. The democratically detrended light curve $\widetilde{D}(t)$ is then defined as

\begin{equation} \label{eq: ensemble median}
    \widetilde{D}(t) = \mathrm{median}_{m=1,...,M} \, \{D_m(t)\}.
\end{equation}

Additionally, in ensemble detrending as implemented in the \texttt{democratic detrender}, we propagate the variance between the different detrending methods. This is done by adding in quadrature for each $i$'th data point the uncertainty on the flux input by the user by the median absolute deviation (MAD) (as implemented in \texttt{scipy.stats.median\_abs\_deviation}) of the different model predictions multiplied by a factor of 1.4826 \citep{RousseeuwCroux1993}. If $\sigma(t)$ is the error on the flux value at time $t$, then the ensemble detrended error, $\widetilde{\sigma}(t)$ would be:

\begin{equation} \label{eq: ensemble errors}
    \widetilde{\sigma}(t) =  \sqrt{ \big[ \sigma(t) \big]^2 +  \big[ 1.4826 * \mathrm{MAD}_{m=1,...,M} \, \{D_m(t)\} \big]^2}.
\end{equation}

We note that the use of the 1.4826 factor assumes that the underlying distribution is approximately Gaussian, and the use of the MAD further assumes that the individual estimates from each $D_m$ are statistically independent. In practice, however, all $D_m$ are detrendings of the same underlying light curve, so some degree of correlation between them is expected. This again underscores the importance of employing detrending methods that differ in their underlying basis sets and cost functions, thereby reducing -- though not eliminating - the potential for shared systematics. Consequently, the expression above should be interpreted as an approximation that assumes Gaussian residuals and uncorrelated samples across the $D_m$.

Thus, in total, $\epsilon(D_M(t), \sigma_M(t))$, the ``ensemble'' function of the \texttt{democratic detrender}, takes a set of M individually detrending light curves and uncertainties and combines them into a single ensemble detrended light curve by both taking the median value at each time, $t$ (see Equation~\ref{eq: ensemble median}), and also inflating the uncertanties by the variance between the surviving M detrending models (see Equation~\ref{eq: ensemble errors}):

\begin{equation} \label{eq: ensemble function}
    \epsilon(D_M(t), \sigma_M(t)) = {\widetilde{D}(t), \widetilde{\sigma}(t)}.
\end{equation}

While the \texttt{democratic detrender} defaults to using the median and MAD when combining detrenders in the ensemble step, we highlight that the user can modify this to mean and standard deviation as the ensemble estimators. Using median and MAD emphasizes robustness to outliers, but it has certain behaviors in the small-$M$ regime: when only two or three detrenders pass the tests, the median simply selects the central value at each timestamp, and the MAD becomes the distance to the nearest neighbor detrender. In such cases one detrending option may contribute little to the final estimate, and crossings between two similar detrenders can lead to small, discrete switches in the median. These effects arise naturally from robust statistics applied to very small samples. Users who prefer smoother behavior may instead select the mean and standard deviation, which are fully supported within the package. The median/MAD pair remains the default because of its resilience to outlying or poorly behaved detrenders, but the ensemble estimator is entirely
user-configurable.

This outlines the fundamental theory behind ensemble detrending (previously also referred to as method marginalized detrending \citep{TeacheyKipping2018} and democratic detrending \citep{Yahalomi2024}) as implemented in the \texttt{democratic detrender}. The \texttt{democratic detrender} provides a flexible framework for detrending exoplanet light curves by combining multiple independent models with different basis sets and cost functions. This ensemble approach focuses on improving accuracy by reducing sensitivity to any single model’s assumptions or biases, while variance inflation accounts for disagreements across methods. By leveraging multiple detrending models, the \texttt{democratic detrender} can better accommodate transient systematics and instrumental effects that may not be fully captured by any one method alone. Its locality flexibility allows different methods to dominate in different parts of the light curve, recognizing that the optimal detrending model for one epoch (or one portion of an epoch) may not be optimal for another. However, this comes at the cost of losing a well-defined analytic detrending model, as the ensemble output is a statistical combination rather than a single functional form. While not suitable as a transit search algorithm, the \texttt{democratic detrender} offers a robust and practical approach that prioritizes minimizing accuracy (rather than precision) by mitigating bias across models.

\section{\texttt{democratic detrender}} \label{sec: detrending methods}

In what follows, we demonstrate the workflow of the \texttt{democratic detrender}, as also shown in Figure~\ref{fig: flow chart}. We will use Kepler-1519\,b as a sample detrending target, demonstrating each step of the \texttt{democratic detrender}. Kepler-1519\,b is a cool gas giant on a $\sim$240 day orbit. It was identified as a target of interest in \citet{KippingYahalomi2023} as it shows short-period transit timing variations (TTVs) that could be consistent with either moon or a companion, non-transiting planet. We highlight that while we use a \textit{Kepler} target as our example, this code has also been extensively tested on \textit{TESS} datasets as well \citep{Kipping2024}.

\subsection{Load Raw Light Curve} \label{sec: load lc}

The first step is loading in the raw light curve. The \texttt{democratic detrender} is setup to load light curves from NASA's Space Telescope Science Institute (STScI) Barbara A. Mikulski Archive for Space Telescopes (MAST)\footnote{\url{https://archive.stsci.edu/}} via the \texttt{lightkurve} package \citep{lightkurve2018}. 

The user must only supply a stellar ID and a planet number. The user need not supply a specific type or format ID, as the \texttt{democratic detrender} queries the SIMBAD Astronomical Database -- CDS (Strasbourg)\footnote{\url{https://simbad.u-strasbg.fr/simbad/}} to find the corresponding TIC ID. If this fails for any reason, the \texttt{democratic detrender} then tries to query NASA's Exoplanet Archive\footnote{\url{https://exoplanetarchive.ipac.caltech.edu/}} to find the relevant TIC ID. For example, for Kepler-1519\,b, the user could supply any of the following IDs, and the \texttt{democratic detrender} would function exactly the same: 2MASS J19464029+4927426, AP J19464029+4927426, Kepler-1519, KIC 11518142, KOI-3762, LAMOST J194640.29+492742.6, TIC 351191596, Gaia DR2 2086851495906130176, and Gaia DR3 2086851495906130176.

With the appropriate TIC ID in hand, the \texttt{democratic detrender} then queries NASA's exoplanet archive in order to load in the planet's period, transit duration, and time of transit midpoint. This information is used when detrending in order to generate the transit mask. The user can optionally supply their own period, transit duration, and/or time of transit midpoint if they so choose.

Next, the \texttt{democratic detrender} loads in the light curve (set of times, flux values, and flux uncertainties) from MAST via the \texttt{lightkurve} package \citep{lightkurve2018}. The user must supply what type of flux data to load, be it Simple Aperture Photometry (SAP) or Pre-search Data Conditioning SAP (PDCSAP). For \textit{TESS} data one can additionally query QLP (Quick Look Pipeline) data. By default, the \texttt{democratic detrender} will load in short cadence data, if it exists, and will only use long cadence data if short cadence observations don't exist for a specific target.

In general, we recommend users to use both SAP and PDCSAP data. This is because PDCSAP data runs the risk of being over-corrected if too many cotrending basis vectors are used. By using both the "raw" SAP data and the PDCSAP data, one can reduce the likelihood of removing signal from the data before fitting for transits or other signals in the time-series photometry.

A transit mask is then created, to mask all transits in the data. By default, we mask $1.1 \, t_\textrm{dur}$ where $t_\textrm{dur}$ is the input transit duration (i.e., a 10\% transit duration buffer). We do this in order to account for possible transit timing variations (TTVs) in the dataset. This transit duration multiplicative factor can be customized to the user's preference and we note for some systems (e.g., those with large TTVs) this will need to be increased.

PDCSAP data is provided to the community already corrected using a contamination factor, called ``CROWDSAP'' in the header, and a flux fraction, called ``FLFRSAP''. CROWDSAP represents the quarterly-averaged ratio of the target flux to the total flux in the photometric aperture and is a scalar value between zero and unity.  FLFRSAP is the fraction of flux from the target star captured by the optimal aperture used in the SAP data and is a scalar between zero and unity. These two factors are then used in correcting the PDCSAP data via the following equation \citep{Stumpe2012, Smith2012, Jenkins2017}: 

\begin{equation} \label{eq: PDCSAP correction}
    \widetilde{y}(t) = \frac{y(t) - (1-c) * \textrm{median}(y(t))}{f},
\end{equation}

where $\widetilde{y}(t)$ is the corrected cotrended flux values, $y(t)$ is the cotrended flux values, $c$ is the CROWDSAP crowding metric, and $f$ is the FLFRSAP flux fraction metric.

Some of this blend data may not include the most recent information, especially with new data from \textit{Gaia}\footnote{\url{https://www.cosmos.esa.int/web/gaia}} \citep{Gaia2016} and the SAP data does not contain this correction. Therefore, by default, we remove this blend correction from the PDCSAP data following \citep{Kipping2010} where $c = 1/\mathcal{B}$ and $\mathcal{B}$ is the blend factor 

\begin{equation}
    \mathcal{B} = \frac{F_* + F_\mathrm{blend} } {F_* }.
\end{equation}

The provided corrected PDCSAP flux values will equal $F_*$ and we want to solve for the $F_\mathrm{total} = F_* + F_\mathrm{blend}$. Thus, we can do so by dividing the provided $F_*$ values by the quarter-averaged CROWDSAP values -- yielding the uncorrected PDCSAP flux values. These uncorrected $y(t)$ values are what we adopt for our PDCSAP data, by default. However, this process of removing the blend factor can be turned off, and the \textit{Kepler} provided corrected PDCSAP flux values can be used. We emphasize that removing the blend factor can meaningfully affect measured transit depths and that we omit this correction by default so that users may (and should) apply their own, up-to-date blend factors based on the most current catalog information.

Finally then, the raw, undetrendred light curve (set of times values, flux values, and flux uncertainties) as well as relevant metadata is returned.

Currently, the \texttt{democratic detrender} is only setup to query and load \textit{Kepler} or \textit{TESS} lightcurves. However, a user can always provide their own raw undetrended LC.

\subsection{Remove Outlier Flux Values} \label{sec: outlier removal}

\begin{figure*}[t!]
    \centering 
    \includegraphics[width=\textwidth]{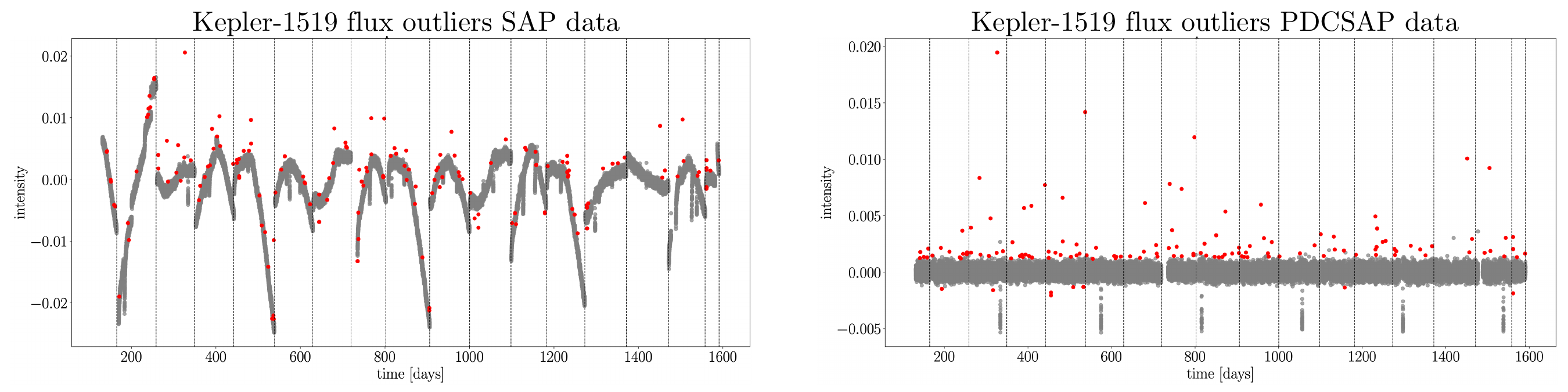}
    \caption{Raw (not detrendend) SAP [left] and PDCSAP [right] light curves for Kepler-1519\,b. Red data points are those that were deemed outliers via moving median rejection and thus removed from subsequent modeling. The grey dotted lines show the different \textit{Kepler} quarters in the data.}
    \label{fig: outlier rejection}
\end{figure*}

Now we begin the process of cleaning the time-series photometry. The first step is to remove outlier flux values. By default, we do so via sigma clipping any flux values in the out-of-transit data that are more than 4$\sigma$ anomalies in a moving median for a time window of 30 cadences (these defaults selections for sigma clipping can also be adjusted by the user). Outlier rejection via sigma clipping can also be turned off. This can be advisable for short cadence \textit{Kepler} datasets or short cadence \textit{TESS} datasets with many sectors of observations, where this process becomes computationally expensive.

\begin{figure*}
    \centering 
    \includegraphics[width=\textwidth]{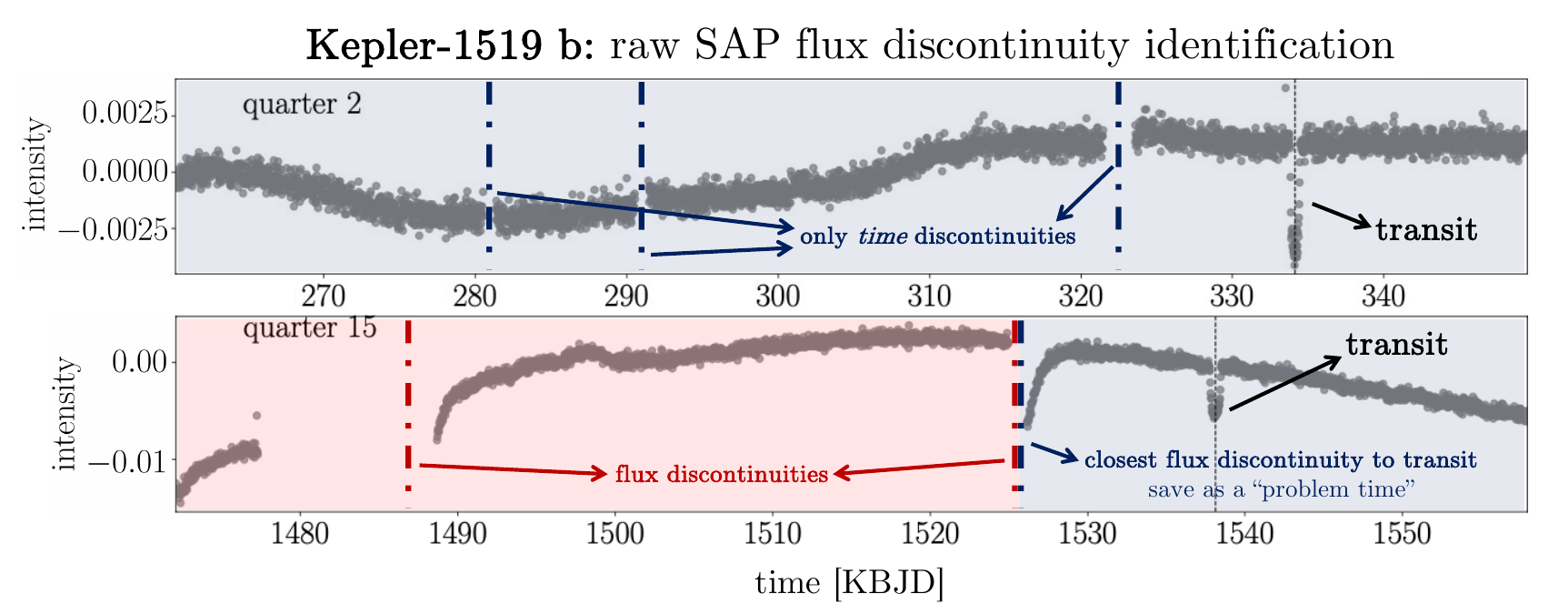}
    \caption{The top row is Kepler-1519\,b's 2$^\textrm{nd}$ quarter of SAP data. In the 2$^\textrm{nd}$ quarter we can see that there are 3 time discontinuities, but as these time gaps don't coincide with a flux intensity discontinuity as well, we wouldn't want to remove any of the data from our detrending. The bottom row is Kepler-1519\,b's 15$^\textrm{th}$ quarter of SAP data. In the 15$^\textrm{th}$ quarter, we can see that there are 2 flux intensity discontinuities that coincide with time discontinuities. In this case, we would want to label the closest flux discontinuity to the transit (KBJD $\sim$ 1527). Once we label this problem time, the \texttt{democratic detrender} will clip the data here, and remove all data from earlier in the quarter, before fitting the detrending models.}
    \label{fig: problem times}
\end{figure*}

Once outlier rejection is completed, Figure~\ref{fig: outlier rejection} is automatically generated. Here the vertical dashed lines represent the quarters, the grey dots represent the non-outlier flux values, and the red dots represent the outlier, and thus removed, values.

\subsection{Identify Times of Flux Discontinuities} \label{sec: problem times}

Detrending models will gain no information on the underlying variability from random variations in the light curve. Further, random (i.e., non-periodic) variations can worsen the accuracy of the detrending models. As explained previously and demonstrated in Figure~\ref{fig: light curve}, the spacecraft, telescope, \& instrument (and even in some cases stellar activity such as stellar flares) can create flux discontinuities. Additionally, these discontinuities can cause detrending models to fail, by forcing the models to try to account for sudden extreme variations. As such, we must remove these discontinuities from the data before detrending. Labeling flux discontinuities is particularly important for using SAP data, where these discontinuities are very prevalent, but it is also good practice for the PDCSAP data. By default, if the user does not label PDCSAP discontinuities, the \texttt{democratic detrender} will assume that there are none. However, if SAP flux discontinuities are identified, the user can choose to use the SAP times of discontinuity for the PDSCAP data as well.

The data is automatically split into quarters (for \textit{Kepler} data) or sectors (for \textit{TESS} data). In between successive quarters or sectors of observation, the target star cannot be expected to fall exactly on the same location in pixel space of the telescope, and so we split the data by quarters or sectors to remove any bias this could introduce into the detrending.

The user must then look through the data and identify flux discontinuities, to be removed from the data. We note that discontinuities only in time space (data gaps that don't correspond with flux jumps) do not need to be labeled as a discontinuity. Identification and labeling of discontinuities can all be done in the \texttt{democratic detrender}, using a built in GUI with sliders that allows the user to identify and save the set of flux discontinuities in the data. An example of the times the user would want to label as discontinuities to be removed from the data can be seen in Figure~\ref{fig: problem times}

\subsection{Detrend Light Curves} \label{sec: detrend}

\begin{figure*}[t!]
    \centering 
    \includegraphics[width=\textwidth]{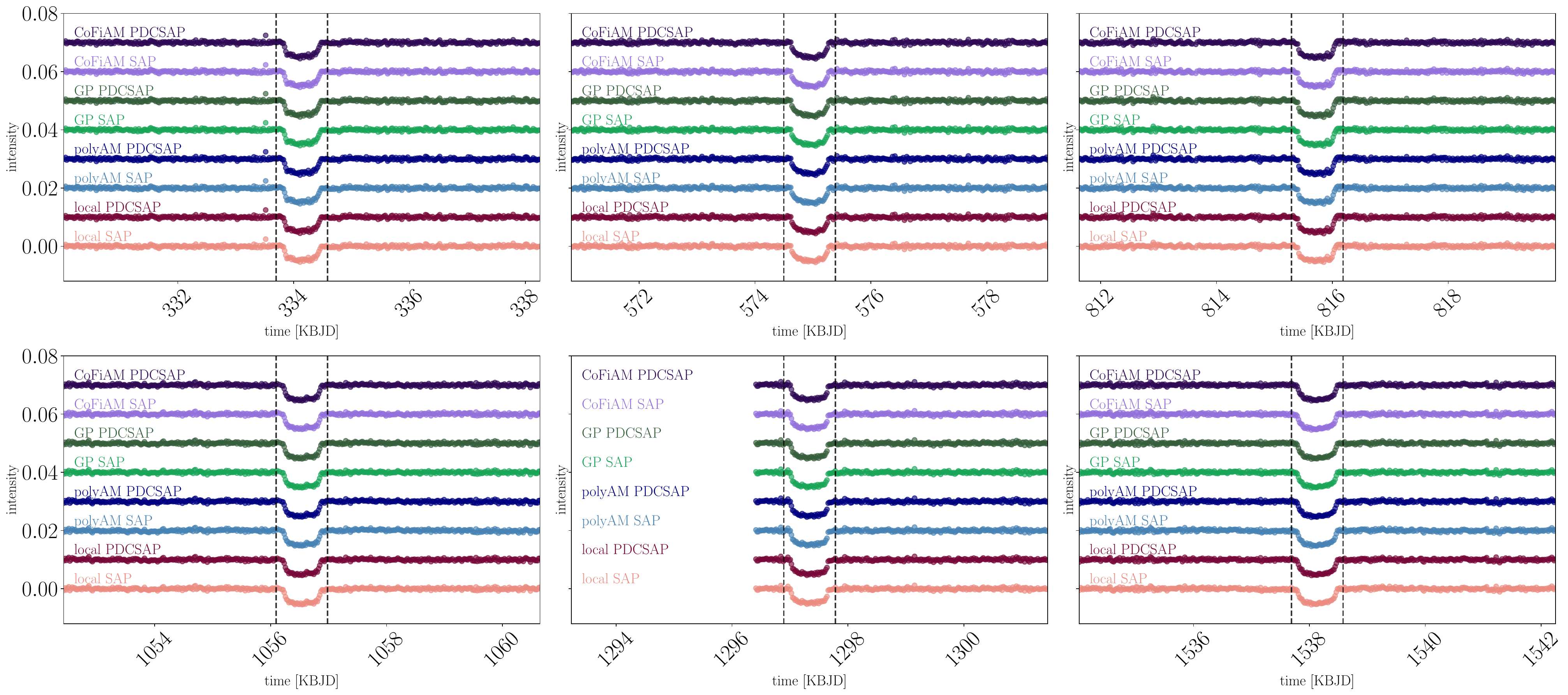}
    \caption{Individually detrended light curves for Kepler-1519\,b. Included in this figure are all 8 default detrending methods in the \texttt{democratic detrender}, which are \texttt{CoFiAM}, \texttt{GP}, \texttt{polyAM}, \texttt{local} run on both the SAP data and PDCSAP data. The vertical dashed lines show the in-transit portion of the light curve (LC) that was ``masked'' and thus not included in the model fitting process. Instead for these portions of the LC, the model fit is extrapolated from the rest of the out-of-transit data such that physical transit features are not removed unintentionally.}
    \label{fig: individually detrended}
\end{figure*}

Finally, with all this data cleaning completed, we can detrend our time-series photometry. As previously stated, there are a number of common detrending used by the astronomy community. By default, the \texttt{democratic detrender} uses four different fitting techniques: (1) \texttt{CoFiAM}, (2) \texttt{GP}, (3) \texttt{polyAM}, and (4) \texttt{local}. These four detrending techniques are applied to both the PDC and SAP time-series photometry, for a total of N=8 individually detrended light curves. An example of these N=8 individually detrended light curves for Kepler-1519\,b can be seen in Figure~\ref{fig: individually detrended}. This figure is also auto-generated by the \texttt{democratic detrender} when detrending completes.

Each of these four detrending models have been used extensively in exoplanet literature, but a brief summary, in part adapted from \citet{Yahalomi2024}, can be seen below:

\begin{itemize}
    
    \item \texttt{CoFiAM}: Cosine Filtering with Autocorrelation Minimisation was first presented in \citet{Kipping2013c} and builds on cosine filtering approach used to study CoRoT data \citep{Mazeh2010}. Specifically, the out-of-transit time-series photometry is regressed over using a high-pass, low-cut filter via a discrete series of harmonic cosine functions, with the functional form given by

    \begin{equation} \label{cofiam}
    f_k(t) = a_0 + \sum_{k=1}^{N_\mathrm{order}} 
    \Big[ x_k \, \sin{\big(\frac{2 \pi t k}{2 D} \big)} + y_k \, \cos{\big(\frac{2 \pi t k}{2 D} \big)}
    \Big] .
    \end{equation}

    Here, D is the total baseline of the time-series data in a given epoch, $t$ is the set of observing times in the epoch, $x_k$ and $y_k$ are free parameters in the model, and $N_\mathrm{order}$ is the highest allowed harmonic order. The selection of $N_\mathrm{order}$ is important, because above a certain limiting threshold, harmonics with a similar timescale as the transit can distort the transit profile.
    In order to determine $N_\mathrm{order}$, we train 30 models (above which numerical instabilites arise), where $N_\mathrm{order}$ ranges from 1 to 30, and at each epoch pick the cosine filter that leads to the least correlated light curve via the Durbin-Watson statistic \citep{DurbinWatson1950, Kipping2013c}.

     \item \texttt{GP}: In this detrending technique, we fit a Gaussian Process to the out-of-transit time-series photometry using the  quasiperiodic kernel presented in \citet{Angus2017} as a reliable model for stellar activity of a rotating star. Specifically, we use a SHOTerm kernel from \texttt{celerite2} via the \texttt{exoplanet} package, which is a stochastically-driven, damped harmonic oscillator \citep[\texttt{exoplanet},][]{Foreman-Mackey2021}, \citep[\texttt{celerite2},][]{celerite1, celerite2}.
    
     \item \texttt{polyAM}: Polynomial detrending with Autocorrelation Minimisation follows a similar process to \texttt{CoFiAM}, except we train 30 models with polynomials as the basis function. Polynomial filtering is a common method for stellar activity detrending \citep[e.g.,][] {Fabrycky2012, Gautier2012, Giles2018}. In \texttt{polyAM}, the 30 different bases models are 1$^\textrm{st}$- to 30$^\textrm{th}$-order polynomials. For each epoch, as in CoFiAM, the least correlated light curve via the Durbin-Watson statistic \citep{DurbinWatson1950} is chosen -- and so only a single polynomial order per epoch is selected.
     
      \item \texttt{local}: The \texttt{local} polynomial method restricts its fit to data within six transit durations of the mid-transit epoch. The \texttt{local} detrending technique again uses 1$^\textrm{st}$- to 30$^\textrm{th}$ order polynomials, but the order of the polynomial is selected via the lowest Bayesian Information Criterion \citep{Schwarz1978} computed -- and so only a single polynomial order per epoch is selected. This a fairly typical detrending method for the analysis of short-period transiters \citep{Sandford2017}.

\end{itemize}

As stated in Section~\ref{sec:intro}, there are many different detrending models used by the exoplanet community to detrend time-series stellar photometry. Therefore, it is likely that there are potential users of the \texttt{democratic detrender} that would like to use a detrending model not provided by the \texttt{democratic detrender}. For example, the Tukey's biweight algorithm, popularized by the \texttt{Wotan} package and \citet{Hippke2019} that demonstrated its utility in modeling stellar time-series photometry is not a detrending model built into the \texttt{democratic detrender}. A tutorial notebook showing how one can include a new detrending model, in this case Tukey's biweight, can be found on the Read the Docs.\footnote{\url{https://democratic-detrender.readthedocs.io/en/latest/}}

Therefore, we have made an optional argument in our code to accept any additional user supplied detrended light curves, to be included in all subsequent steps. Additionally, the user can decide which of the 4 included detrending models to use in their detrending. 

We again emphasize the importance of choosing detrending models that use different basis sets and different cost functions. One might notice that of the included detrending models, \texttt{CoFiAM} and \texttt{polyAM} both select the order of the basis function via autocorrelation minimization (i.e., the same cost function), but the basis functions in these two methods are quite different. Additionally, \texttt{polyAM} and \texttt{local} both fit polynomials to the LCs, however, they are trained on different subsets of the data and the cost function that is used to determine the order of the selected polynomial is different.

\subsection{Model Rejection} \label{sec: model rejection}

We now have by default N=8 individually detrended light curves produced by the \texttt{democratic detrender} and any user supplied individually detrended light curves. The SAP datasets, PDCSAP datasets, and user supplied detrended LCs could have the same or slightly different time arrays in the datasets. This is because it is possible that with discontinuity identification certain portions of the dataset are removed from SAP data, but not from PDCSAP data, for example. So before we can aggregate these light curves together via the ensemble step, we must make them all have the same exact time values. In order to do so, we pad the flux and flux uncertainty sets with a \textit{NaN} value for any time that is missing in a given individually detrended dataset.

For each of the $N$ individually detrended light curves, we want to test the Gaussianity of each detrending model for each epoch. To this end, we perform two bootstrapping tests via Monte Carlo simulations: (1) a test of the autocorrelation of the detrended flux values via the Durbin-Watson (DW) statistic and (2) a test of excess noise in the light curve via the binning vs. root-mean-square variance expected behavior. This allows us to remove non-Gaussian and autocorrelated detrended LCs before the ensemble step.

\begin{enumerate}
    \item \textbf{Autocorrelation test via DW metric}:

    \begin{figure}
        \centering 
        \includegraphics[width=\columnwidth]{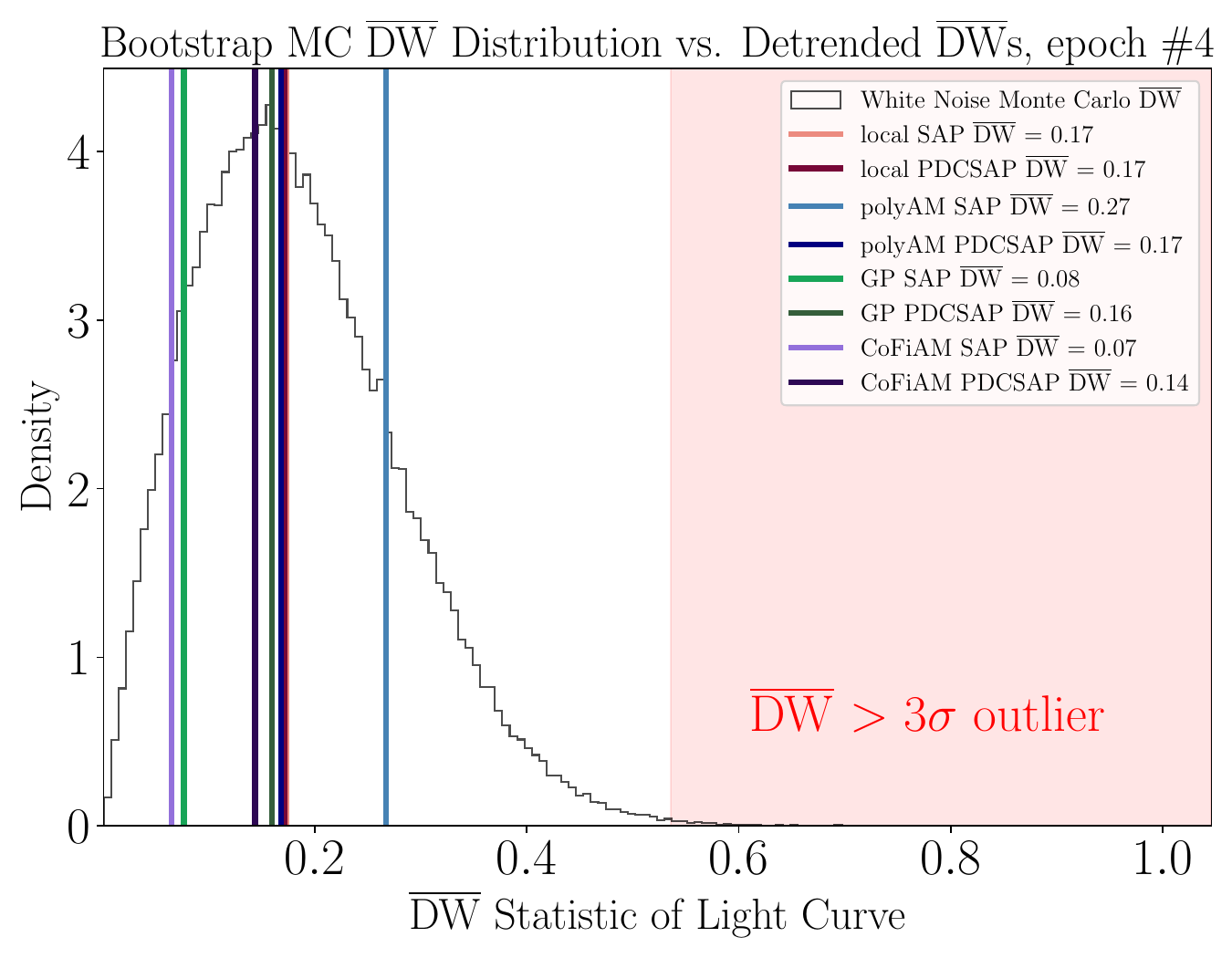}
        \caption{Bootstrapping test of autocorrelation via the DW metric for Kepler-1519\,b's $4^\textrm{th}$ transit epoch. The histogram shows the Monte Carlo results for a slightly modified DW metric, $\overline{DW}$, as defined by Equation~\ref{eq: dw_mod} for simulated white noise with the same uncertainties as the light curve. The vertical lines show the $\overline{DW}$ values for each of the 8 detrending models. Shaded in red are $\overline{DW}$ values that are more than 3-$\sigma$ outliers. Any detrending model that falls in this shaded region is excluded from the final ensemble detrended model.}
        \label{fig: dw test}
    \end{figure}

    The Durbin-Watson (DW) statistic, as first presented in \citet{DurbinWatson1950}, is a metric used that can be used to detect autocorrelation, or the correlation of a signal with a delayed copy of itself, in time-series analysis. In our context, DW is defined as

    \begin{equation} \label{eq: dw}
        DW = \frac{\sum_{t=2}^{T} (e_t - e_{t-1})^2}{\sum_{t=1}^{T} e_t^2},
    \end{equation}

    where $e_t$ is the residual at time $t$. The value of the DW statistic ranges from 0 to 4, where DW of 2 means no autocorrelation.

    As implemented here (as can be seen by the $e_{t-1}$ term in Equation \ref{eq: dw}), DW is the lag one autocorrelation with the timescale of the cadence. In order to use a different timescale (e.g., an hour) one would have to first bin the time series to that cadence.

    If the data was infinitely long, then we would expect the DW statistic of a detrended dataset to approach 2. As our dataset is finite, instead we can perform a bootstrap test of the DW statistic for each individually detrended LC at each epoch by creating a Monte Carlo (MC) simulation with 100,000 realizations of random Gaussian noise with a standard deviation equal to the provided flux uncertanties. 

    We then determine the DW statistic for each of the 100,000 MC realizations using Equation~\ref{eq: dw}. As there is a significant datagap at each transit, we determine the DW statistic for the pre-transit and post-transit data separately and then combine them by adding in slightly modified quadrature as follows
    
    \begin{equation} \label{eq: dw_mod}
        \overline{DW} = \sqrt{(DW_\textrm{pre}-2)^2 + (DW_\textrm{post}-2)^2}.
    \end{equation} 
    
    For the $\overline{DW}$ statistic, a dataset with no autocorrelation will have a value of 0.

    We can then use the inverse error function to determine the location of the one-sided 3-$\sigma$ upper bound limit by identifying the 99.7th percentile of the $\overline{DW}$ statistic from the Monte Carlo simulations. Then we determine the $\overline{DW}$ statistic for each individually detrended light curve at each epoch, and remove any individually detrended epochs that are more than 3-$\sigma$ outliers. As $\overline{DW}$ will never be less than zero and a $\overline{DW}$ value closer to zero is indicative of a dataset that shows less autocorrelation, there is only a 3-$\sigma$ upper bound. See Figure~\ref{fig: dw test} for an example of the autocorrelation via DW test for a sample epoch of Kepler-1519\,b. The sigma cutoff parameter is customizable (e.g., users can change the cutoff to be a stricter 2-$\sigma$ upper bound) if they want to be more or less strict on what is considered an outlier.

    \item \textbf{Binning vs. RMS test}: 

    \begin{figure}
        \centering 
        \includegraphics[width=\columnwidth]{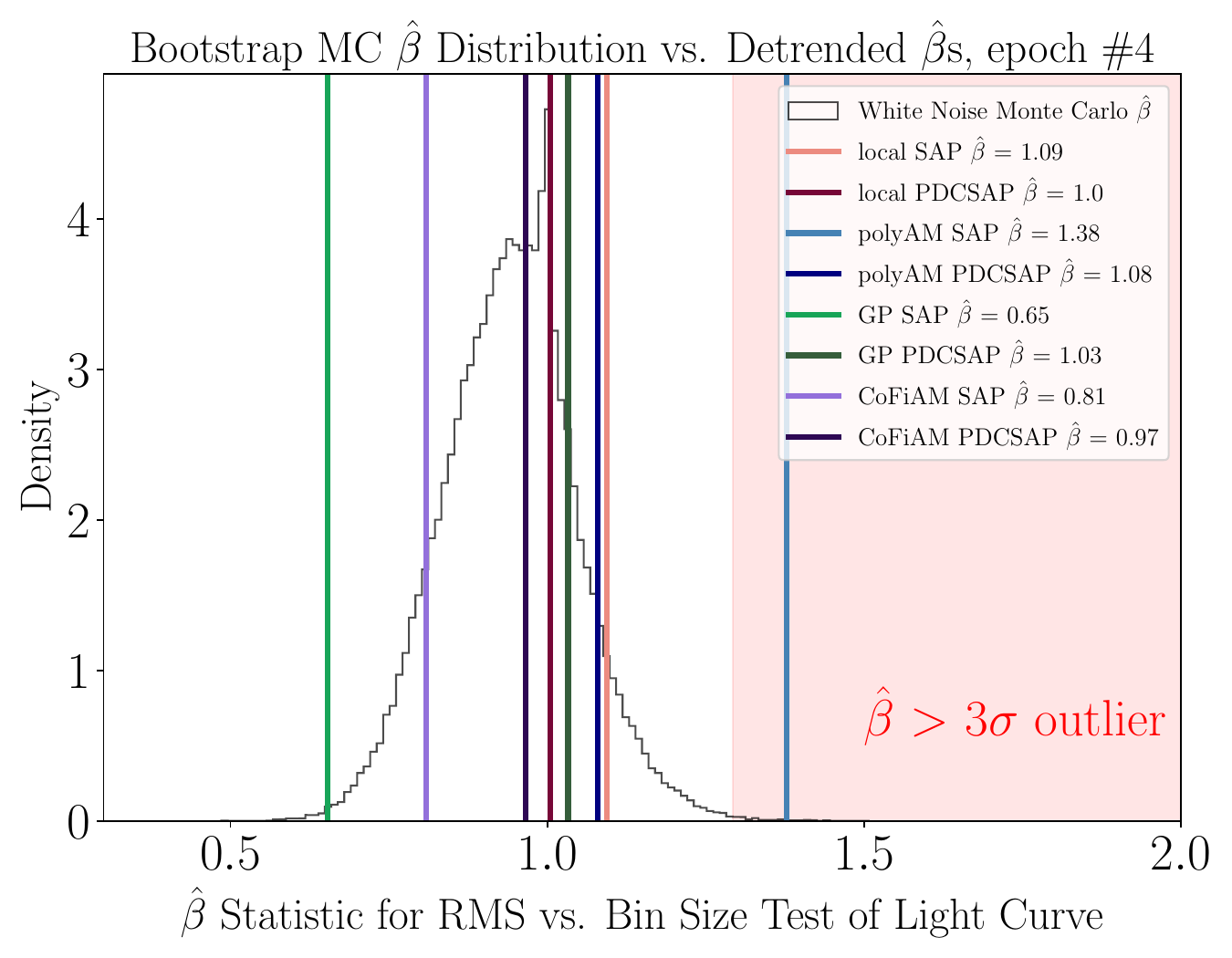}
        \caption{Bootstrapping test of root-mean-square (RMS) vs. bin for Kepler-1519\,b's $4^\textrm{th}$ transit epoch. The histogram show the Monte Carlo $\hat{\beta}$ values (see Equation~\ref{eq: binning test} and the subsequent definition of $\hat{\beta}$). A larger $\hat{\beta}$ value indicates the presence of coherent (red) noise. The vertical lines show the $\hat{\beta}$ values for each of the 8 detrending models. Shaded in red are $\hat{\beta}$ values that are more than 3-$\sigma$ outliers. Any detrending model that falls in this shaded region would be excluded from the final ensemble detrended model.}
        \label{fig: binning test}
    \end{figure}

    We also run a ``time-averaging'' test of correlated red noise that was first propounded in \citet{Pont2006} and we follow closely to the formalism described in detail in \citet{CarterWinn2009}. In short, we determine the standard deviation of both the unbinned data, $\hat{\sigma}_1$, and also the time averaged data, $\hat{\sigma}_n$ where every $n$ points have been averaged (thus creating m time bins.) In the absence of correlated (red) noise, we expect white noise will behave as

    \begin{equation} \label{eq: binning test}
        \hat{\sigma}_n = \frac{\hat{\sigma}_1}{n^{1/2}} \, \Big( \frac{m} {m - 1} \Big )^{1/2}.
    \end{equation}

    In the \texttt{democratic detrender}, we use a minimum ${M_\textrm{bins}}$ value of 1 data point and a maximum ${M_\textrm{bins}}$ value equal to 1/10 the length of the total included dataset. For computational efficiency, we bin by index rather than time, and assume that the data near transit is approximately uniformly spaced. As this is a bootstrap test, if this is not the case, the effect should be mitigated as we bin the same way for both the MC simulated data and the real detrending model data.
    
    In the presence of correlated noise, we expect that there will be excess noise above the theoretical expression in Equation~\ref{eq: binning test}. We define a factor, $\hat{\beta}_n$, that represents the ratio between the observed standard deviation of the binned data, $\hat{\sigma}'_n$ and its white noise theoretical expression ($\hat{\beta}_n = \hat{\sigma}'_n / \hat{\sigma}_n$). We then define the estimator, $\hat{\beta}$, defined as the median of all $\hat{\beta}_n$ values, which represents the amount of excess red noise in the data. A value of $\hat{\beta}$ that greatly exceeds 1 is indicative of a coherent red noise.

    We start our bootstrap test by simulating 100,000 light curves for each epoch with random Gaussian noise with the provided flux uncertainties. We determine the $\hat{\beta}$ value for each of these Monte Carlo simulations. For this set of $\hat{\beta}$ values, we then determine the 3-$\sigma$ boundary limit for each epoch using the inverse error function, as described above.

    Finally, for each epoch of each individually detrended light curve, we calculate the $\hat{\beta}$ value and remove any individually detrended light curves $\hat{\beta}$ values that are larger than the 3-$\sigma$ cutoff. A $\hat{\beta}$ that is so small that it is a 3-$\sigma$ outlier in the other direction is indicative that the errors are likely overestimated. As this isn't indicative of coherent noise, we don't remove detrending models with this behavior. See Figure~\ref{fig: binning test} for an example of the binning vs RMS test for a sample epoch of Kepler-1519\,b. The sigma cutoff parameter is customizable (e.g., users can change the cutoff to be a stricter 2-$\sigma$ upper bound) if they want to be more or less strict on what is considered an outlier.

\end{enumerate}

We adopt a 3-$\sigma$ threshold for these tests as a pragmatic balance between sensitivity and robustness. Our aim is to identify and down–weight detrendings that exhibit strongly non-Gaussian or red-noise–dominated behavior, rather than to reject mildly deviant but still acceptable realizations. A 3-$\sigma$ cut provides an effective way to flag detrenders whose residual structure is inconsistent with the white-noise assumptions underlying the ensemble, while avoiding overly aggressive pruning.

As shown in Figure~\ref{fig: dw test} and Figure~\ref{fig: binning test}, for the 4$\mathrm{th}$ epoch of Kepler-1519\,b, 8 of the 8 detrending methods pass the DW autocorrelation test and 7 of the 8 detrending methods pass the binning vs. RMS test. Therefore, for the 4$\mathrm{th}$ epoch of Kepler-1519\,b, we have M=7 detrending models to pass into the ensemble function, and we will not include \texttt{polyAM} SAP in our ensemble function for the 4th epoch.

\begin{figure*}
    \centering 
    \includegraphics[width=\textwidth]{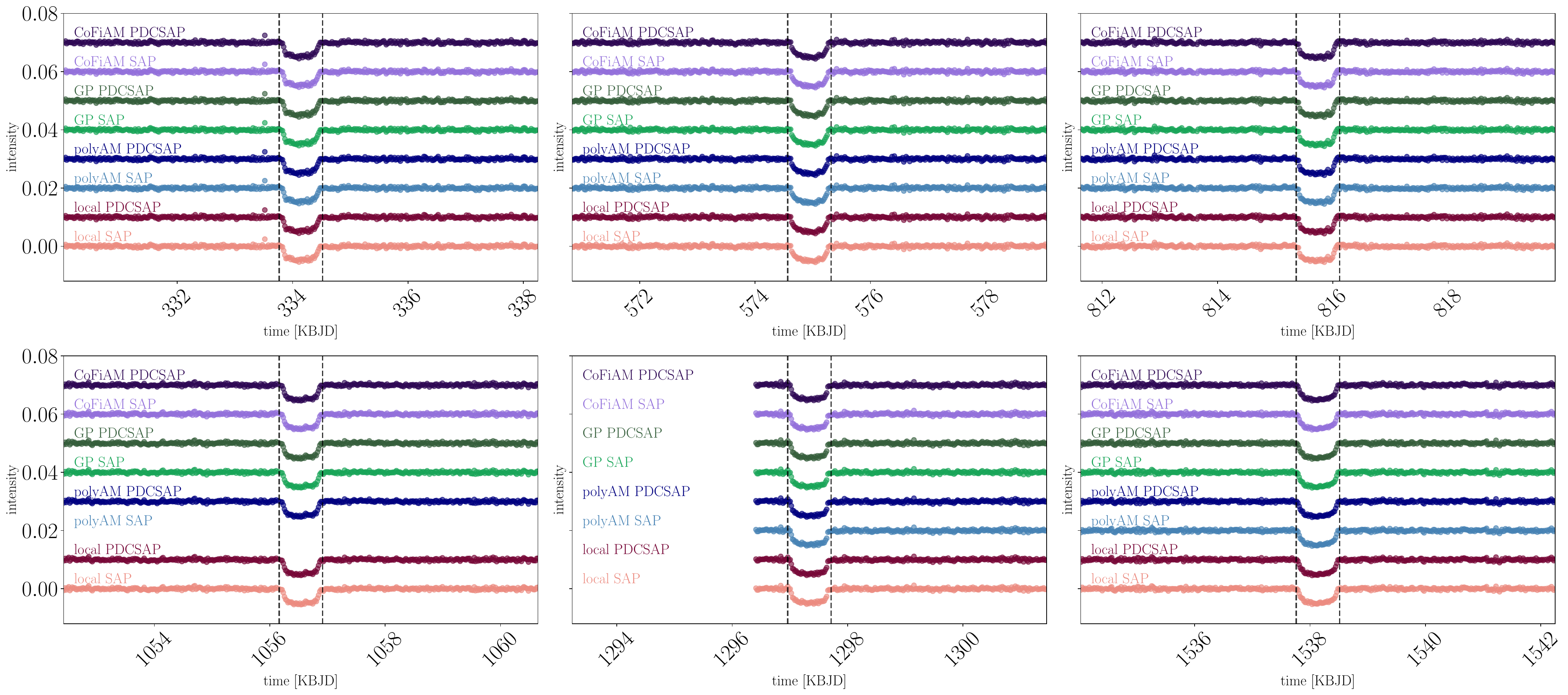}
    \caption{Individually detrended light curves for Kepler-1519\,b that passed both tests of correlation. Included in this figure are all 8 default detrending methods in the \texttt{democratic detrender}, which are \texttt{CoFiAM}, \texttt{GP}, \texttt{polyAM}, \texttt{local} run on both the SAP data and PDCSAP data. We can see that for the 4$^\mathrm{th}$ epoch, one of the detrending models failed at least one of the tests while for the other 5 epochs, all of the detrending models passed both tests. The vertical dashed lines show the in-transit portion of the light curve (LC) that was ``masked'' and thus not included in the model fitting process. Instead for these portions of the LC, the model fit is extrapolated from the rest of the out-of-transit data such that physical transit features are not removed unintentionally.}
    \label{fig: individually detrended post rejection}
\end{figure*}

\begin{figure*}
    \centering 
    \includegraphics[width=\textwidth]{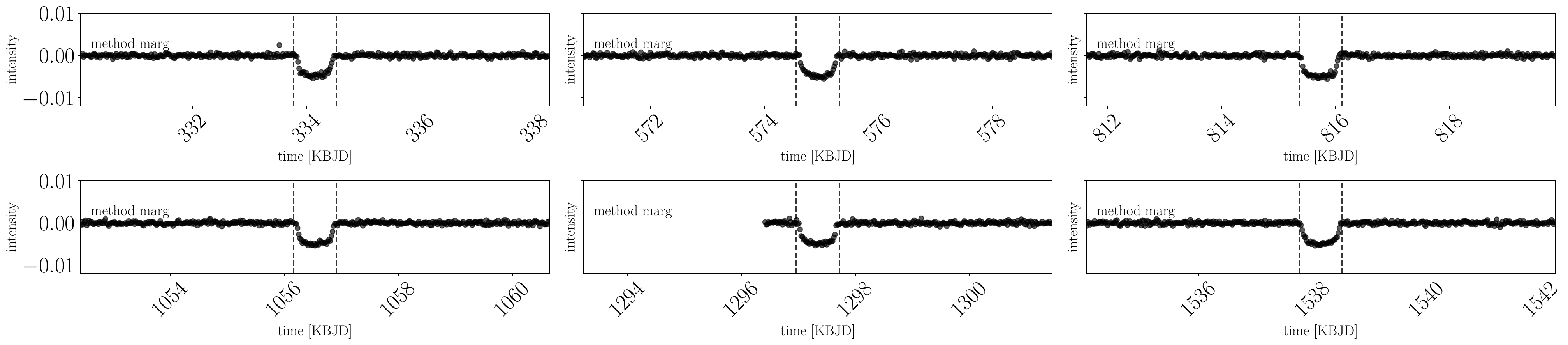}
    \caption{Final ensemble detrended light curves for Kepler-1519\,b. Included in this ensemble detrending are are all 8 default detrending methods that passed both bootstrapping tests of correlated noise in the \texttt{democratic detrender}, which are \texttt{CoFiAM}, \texttt{GP}, \texttt{polyAM}, \texttt{local} run on both the SAP data and PDCSAP data. See Figure~\ref{fig: individually detrended post rejection} for which detrending models passed the tests for each epoch. The vertical dashed lines show the in-transit portion of the light curve (LC) that was ``masked'' and thus not included in the model fitting process. Instead for these portions of the LC, the model fit is extrapolated from the rest of the out-of-transit data such that physical transit features are not removed unintentionally.}
    \label{fig: ensemble detrended}
\end{figure*}

For each epoch, there can be a different number of $M$ detrending models that survive the detrending tests. Figure~\ref{fig: individually detrended post rejection} shows the ensemble of individually detrended light curves for Kepler-1519\,b that pass both tests. Here we can see that by including $N > 1$ detrending models in our whitening step, we allow for robust tests of whiteness, as we down select to only the $M \leq N$ white detrended light curves.

\subsection{Ensemble Step} \label{sec: ensemble step}

Finally, now, we can perform the ensemble step as presented in Equation~\ref{eq: ensemble function}. We take the median value at each time step for the $M$ surviving individually detrended light curves at each epoch, as described by Equation~\ref{eq: ensemble median}. Additionally, we inflate the uncertainties by the variance between the different $M$ surviving models, as described by Equation~\ref{eq: ensemble errors}. This is described in depth in Section~\ref{sec: ensemble detrending}. Doing so results in our final ensemble detrended light curve, as shown in Figure~\ref{fig: ensemble detrended}. 

As a final test, we ensure that, for each epoch, the ensemble detrended light curve does not exhibit signs of correlated noise by again using the same 2 tests of Gaussianity and uncorrelated noise, described in Section~\ref{sec: model rejection}.

\section{Conclusion}

Space-based time-series stellar photometry includes many different contributing factors. These include detector noise, spacecraft, instrument, and telescope induced noise, stellar activity, and eclipses. In modeling eclipses (or planetary transits) it is essential to efficiently remove unwanted activity from these time-series datasets both accurately and precisely and without introducing additional non-physical subtle features in the data. There exists no perfect model for this unwanted activity, and the many different detrending models used by the exoplanet community each have their own strengths and weaknesses \citep{Hippke2019}.

For modeling transits, space-based time-series will frequently have a baseline of out-of-transit data that greatly exceeds the in-transit data baseline. When this is the case, it is an appropriate approximation to model the transit in a two-stepped approach, first removing unwanted activity and then fitting an exoplanet transit to the detrended data.

When modeling such data where implementing this two-stepped approach is appropriate, in this manuscript we present a new way to mitigate the imperfections of detrending models by detrending via an ensemble-based approach with a community-of-models. We additionally present a set of tests of each individually detrended light curves for correlated noise, ensuring that any detrending model used in the final ensemble detrended light curve is consistent with the expected Gaussian uncorrelated white noise properties.

By doing so, we can reduce the dependency of our final ensemble detrended light curve on the assumptions of different detrending models -- therefore improving our detrended light curve by mitigating model dependency and propagating information regarding uncertainties between the individual detrending models. This can aid in more precise transit modeling and will reduce the likelihood of erroneous non-physical features being propagated into our fitted planetary parameters.

Finally, we present an open-source, modular, and scalable coding package, the \texttt{democratic detrender}, that can perform out-of-the-box ensemble detrending on all \textit{Kepler}, \textit{TESS}, and any user-supplied time-series light curve.

\begin{acknowledgments}

We thank the anonymous reviewer who’s feedback improved this manuscript. D.A.Y. and D.K. acknowledge support from NASA Grant \#80NSSC21K0960. D.A.Y. acknowledges support from the NASA/NY Space Grant. D.A.Y. is supported by Flatiron Research Fellowships at the Flatiron Institute, a division of the Simons Foundation. D.A.Y. thanks the LSST-DA Data Science Fellowship Program, which is funded by LSST-DA, the Brinson Foundation, and the Moore Foundation; his participation in the program has benefited this work. J.D.T. acknowledges funding support for this research by the TESS Guest Investigator Program G06165. 

Some of the code used in the \texttt{democratic detrender} was adopted from the \texttt{MoonPy} package \citep{Teachey2021}.

This research made use of \textsf{exoplanet} \citep{exoplanet:joss, exoplanet:zenodo} and its dependencies \citep{exoplanet:arviz, exoplanet:astropy13, exoplanet:astropy18, exoplanet:kipping13b, exoplanet:pymc3, exoplanet:theano}.

D.A.Y and D.K. thank the following for their generous support to the Cool Worlds Lab:
Douglas Daughaday,
Elena West,
Tristan Zajonc,
Alex de Vaal,
Mark Elliott,
Stephen Lee,
Zachary Danielson,
Chad Souter,
Marcus Gillette,
Tina Jeffcoat,
Jason Rockett,
Tom Donkin,
Andrew Schoen,
Reza Ramezankhani,
Steven Marks,
Nicholas Gebben,
Mike Hedlund,
Leigh Deacon,
Ryan Provost,
Nicholas De Haan,
Emerson Garland,
The Queen Road Foundation Inc,
Scott Thayer,
Frank Blood,
Ieuan Williams,
Xinyu Yao,
Axel Nimmerjahn,
Brian Cartmell,
\&
Guillaume Le Saint.

\end{acknowledgments}

%


\software{
\texttt{exoplanet} \citep{Foreman-Mackey2021}, $\,$
\texttt{matplotlib} \citep{matplotlib}, $\,$
\texttt{numpy} \citep{numpy}, $\,$
\texttt{scipy} \citep{scipy}$\,$
\texttt{PyMC3} \citep{Salvatier2016} $\,$
\texttt{corner} \citep{Foreman-Mackey2016} $\,$
\texttt{ChatGPT} was utilized to improve wording at the sentence level and assist with coding -- last accessed in 2025 November.}


\bibliography{main}{}

@ARTICLE{Gaia2016,
       author = {{Gaia Collaboration} and {Prusti}, T. and {de Bruijne}, J.~H.~J. and {Brown}, A.~G.~A. and {Vallenari}, A. and {Babusiaux}, C. and {Bailer-Jones}, C.~A.~L. and {Bastian}, U. and {Biermann}, M. and {Evans}, D.~W. and {Eyer}, L. and {Jansen}, F. and {Jordi}, C. and {Klioner}, S.~A. and {Lammers}, U. and {Lindegren}, L. and {Luri}, X. and {Mignard}, F. and {Milligan}, D.~J. and {Panem}, C. and {Poinsignon}, V. and {Pourbaix}, D. and {Randich}, S. and {Sarri}, G. and {Sartoretti}, P. and {Siddiqui}, H.~I. and {Soubiran}, C. and {Valette}, V. and {van Leeuwen}, F. and {Walton}, N.~A. and {Aerts}, C. and {Arenou}, F. and {Cropper}, M. and {Drimmel}, R. and {H{\o}g}, E. and {Katz}, D. and {Lattanzi}, M.~G. and {O'Mullane}, W. and {Grebel}, E.~K. and {Holland}, A.~D. and {Huc}, C. and {Passot}, X. and {Bramante}, L. and {Cacciari}, C. and {Casta{\~n}eda}, J. and {Chaoul}, L. and {Cheek}, N. and {De Angeli}, F. and {Fabricius}, C. and {Guerra}, R. and {Hern{\'a}ndez}, J. and {Jean-Antoine-Piccolo}, A. and {Masana}, E. and {Messineo}, R. and {Mowlavi}, N. and {Nienartowicz}, K. and {Ord{\'o}{\~n}ez-Blanco}, D. and {Panuzzo}, P. and {Portell}, J. and {Richards}, P.~J. and {Riello}, M. and {Seabroke}, G.~M. and {Tanga}, P. and {Th{\'e}venin}, F. and {Torra}, J. and {Els}, S.~G. and {Gracia-Abril}, G. and {Comoretto}, G. and {Garcia-Reinaldos}, M. and {Lock}, T. and {Mercier}, E. and {Altmann}, M. and {Andrae}, R. and {Astraatmadja}, T.~L. and {Bellas-Velidis}, I. and {Benson}, K. and {Berthier}, J. and {Blomme}, R. and {Busso}, G. and {Carry}, B. and {Cellino}, A. and {Clementini}, G. and {Cowell}, S. and {Creevey}, O. and {Cuypers}, J. and {Davidson}, M. and {De Ridder}, J. and {de Torres}, A. and {Delchambre}, L. and {Dell'Oro}, A. and {Ducourant}, C. and {Fr{\'e}mat}, Y. and {Garc{\'\i}a-Torres}, M. and {Gosset}, E. and {Halbwachs}, J. -L. and {Hambly}, N.~C. and {Harrison}, D.~L. and {Hauser}, M. and {Hestroffer}, D. and {Hodgkin}, S.~T. and {Huckle}, H.~E. and {Hutton}, A. and {Jasniewicz}, G. and {Jordan}, S. and {Kontizas}, M. and {Korn}, A.~J. and {Lanzafame}, A.~C. and {Manteiga}, M. and {Moitinho}, A. and {Muinonen}, K. and {Osinde}, J. and {Pancino}, E. and {Pauwels}, T. and {Petit}, J. -M. and {Recio-Blanco}, A. and {Robin}, A.~C. and {Sarro}, L.~M. and {Siopis}, C. and {Smith}, M. and {Smith}, K.~W. and {Sozzetti}, A. and {Thuillot}, W. and {van Reeven}, W. and {Viala}, Y. and {Abbas}, U. and {Abreu Aramburu}, A. and {Accart}, S. and {Aguado}, J.~J. and {Allan}, P.~M. and {Allasia}, W. and {Altavilla}, G. and {{\'A}lvarez}, M.~A. and {Alves}, J. and {Anderson}, R.~I. and {Andrei}, A.~H. and {Anglada Varela}, E. and {Antiche}, E. and {Antoja}, T. and {Ant{\'o}n}, S. and {Arcay}, B. and {Atzei}, A. and {Ayache}, L. and {Bach}, N. and {Baker}, S.~G. and {Balaguer-N{\'u}{\~n}ez}, L. and {Barache}, C. and {Barata}, C. and {Barbier}, A. and {Barblan}, F. and {Baroni}, M. and {Barrado y Navascu{\'e}s}, D. and {Barros}, M. and {Barstow}, M.~A. and {Becciani}, U. and {Bellazzini}, M. and {Bellei}, G. and {Bello Garc{\'\i}a}, A. and {Belokurov}, V. and {Bendjoya}, P. and {Berihuete}, A. and {Bianchi}, L. and {Bienaym{\'e}}, O. and {Billebaud}, F. and {Blagorodnova}, N. and {Blanco-Cuaresma}, S. and {Boch}, T. and {Bombrun}, A. and {Borrachero}, R. and {Bouquillon}, S. and {Bourda}, G. and {Bouy}, H. and {Bragaglia}, A. and {Breddels}, M.~A. and {Brouillet}, N. and {Br{\"u}semeister}, T. and {Bucciarelli}, B. and {Budnik}, F. and {Burgess}, P. and {Burgon}, R. and {Burlacu}, A. and {Busonero}, D. and {Buzzi}, R. and {Caffau}, E. and {Cambras}, J. and {Campbell}, H. and {Cancelliere}, R. and {Cantat-Gaudin}, T. and {Carlucci}, T. and {Carrasco}, J.~M. and {Castellani}, M. and {Charlot}, P. and {Charnas}, J. and {Charvet}, P. and {Chassat}, F. and {Chiavassa}, A. and {Clotet}, M. and {Cocozza}, G. and {Collins}, R.~S. and {Collins}, P. and {Costigan}, G. and {Crifo}, F. and {Cross}, N.~J.~G. and {Crosta}, M. and {Crowley}, C. and {Dafonte}, C. and {Damerdji}, Y. and {Dapergolas}, A. and {David}, P. and {David}, M. and {De Cat}, P. and {de Felice}, F. and {de Laverny}, P. and {De Luise}, F. and {De March}, R. and {de Martino}, D. and {de Souza}, R. and {Debosscher}, J. and {del Pozo}, E. and {Delbo}, M. and {Delgado}, A. and {Delgado}, H.~E. and {di Marco}, F. and {Di Matteo}, P. and {Diakite}, S. and {Distefano}, E. and {Dolding}, C. and {Dos Anjos}, S. and {Drazinos}, P. and {Dur{\'a}n}, J. and {Dzigan}, Y. and {Ecale}, E. and {Edvardsson}, B. and {Enke}, H. and {Erdmann}, M. and {Escolar}, D. and {Espina}, M. and {Evans}, N.~W. and {Eynard Bontemps}, G. and {Fabre}, C. and {Fabrizio}, M. and {Faigler}, S. and {Falc{\~a}o}, A.~J. and {Farr{\`a}s Casas}, M. and {Faye}, F. and {Federici}, L. and {Fedorets}, G. and {Fern{\'a}ndez-Hern{\'a}ndez}, J. and {Fernique}, P. and {Fienga}, A. and {Figueras}, F. and {Filippi}, F. and {Findeisen}, K. and {Fonti}, A. and {Fouesneau}, M. and {Fraile}, E. and {Fraser}, M. and {Fuchs}, J. and {Furnell}, R. and {Gai}, M. and {Galleti}, S. and {Galluccio}, L. and {Garabato}, D. and {Garc{\'\i}a-Sedano}, F. and {Gar{\'e}}, P. and {Garofalo}, A. and {Garralda}, N. and {Gavras}, P. and {Gerssen}, J. and {Geyer}, R. and {Gilmore}, G. and {Girona}, S. and {Giuffrida}, G. and {Gomes}, M. and {Gonz{\'a}lez-Marcos}, A. and {Gonz{\'a}lez-N{\'u}{\~n}ez}, J. and {Gonz{\'a}lez-Vidal}, J.~J. and {Granvik}, M. and {Guerrier}, A. and {Guillout}, P. and {Guiraud}, J. and {G{\'u}rpide}, A. and {Guti{\'e}rrez-S{\'a}nchez}, R. and {Guy}, L.~P. and {Haigron}, R. and {Hatzidimitriou}, D. and {Haywood}, M. and {Heiter}, U. and {Helmi}, A. and {Hobbs}, D. and {Hofmann}, W. and {Holl}, B. and {Holland}, G. and {Hunt}, J.~A.~S. and {Hypki}, A. and {Icardi}, V. and {Irwin}, M. and {Jevardat de Fombelle}, G. and {Jofr{\'e}}, P. and {Jonker}, P.~G. and {Jorissen}, A. and {Julbe}, F. and {Karampelas}, A. and {Kochoska}, A. and {Kohley}, R. and {Kolenberg}, K. and {Kontizas}, E. and {Koposov}, S.~E. and {Kordopatis}, G. and {Koubsky}, P. and {Kowalczyk}, A. and {Krone-Martins}, A. and {Kudryashova}, M. and {Kull}, I. and {Bachchan}, R.~K. and {Lacoste-Seris}, F. and {Lanza}, A.~F. and {Lavigne}, J. -B. and {Le Poncin-Lafitte}, C. and {Lebreton}, Y. and {Lebzelter}, T. and {Leccia}, S. and {Leclerc}, N. and {Lecoeur-Taibi}, I. and {Lemaitre}, V. and {Lenhardt}, H. and {Leroux}, F. and {Liao}, S. and {Licata}, E. and {Lindstr{\o}m}, H.~E.~P. and {Lister}, T.~A. and {Livanou}, E. and {Lobel}, A. and {L{\"o}ffler}, W. and {L{\'o}pez}, M. and {Lopez-Lozano}, A. and {Lorenz}, D. and {Loureiro}, T. and {MacDonald}, I. and {Magalh{\~a}es Fernandes}, T. and {Managau}, S. and {Mann}, R.~G. and {Mantelet}, G. and {Marchal}, O. and {Marchant}, J.~M. and {Marconi}, M. and {Marie}, J. and {Marinoni}, S. and {Marrese}, P.~M. and {Marschalk{\'o}}, G. and {Marshall}, D.~J. and {Mart{\'\i}n-Fleitas}, J.~M. and {Martino}, M. and {Mary}, N. and {Matijevi{\v{c}}}, G. and {Mazeh}, T. and {McMillan}, P.~J. and {Messina}, S. and {Mestre}, A. and {Michalik}, D. and {Millar}, N.~R. and {Miranda}, B.~M.~H. and {Molina}, D. and {Molinaro}, R. and {Molinaro}, M. and {Moln{\'a}r}, L. and {Moniez}, M. and {Montegriffo}, P. and {Monteiro}, D. and {Mor}, R. and {Mora}, A. and {Morbidelli}, R. and {Morel}, T. and {Morgenthaler}, S. and {Morley}, T. and {Morris}, D. and {Mulone}, A.~F. and {Muraveva}, T. and {Musella}, I. and {Narbonne}, J. and {Nelemans}, G. and {Nicastro}, L. and {Noval}, L. and {Ord{\'e}novic}, C. and {Ordieres-Mer{\'e}}, J. and {Osborne}, P. and {Pagani}, C. and {Pagano}, I. and {Pailler}, F. and {Palacin}, H. and {Palaversa}, L. and {Parsons}, P. and {Paulsen}, T. and {Pecoraro}, M. and {Pedrosa}, R. and {Pentik{\"a}inen}, H. and {Pereira}, J. and {Pichon}, B. and {Piersimoni}, A.~M. and {Pineau}, F. -X. and {Plachy}, E. and {Plum}, G. and {Poujoulet}, E. and {Pr{\v{s}}a}, A. and {Pulone}, L. and {Ragaini}, S. and {Rago}, S. and {Rambaux}, N. and {Ramos-Lerate}, M. and {Ranalli}, P. and {Rauw}, G. and {Read}, A. and {Regibo}, S. and {Renk}, F. and {Reyl{\'e}}, C. and {Ribeiro}, R.~A. and {Rimoldini}, L. and {Ripepi}, V. and {Riva}, A. and {Rixon}, G. and {Roelens}, M. and {Romero-G{\'o}mez}, M. and {Rowell}, N. and {Royer}, F. and {Rudolph}, A. and {Ruiz-Dern}, L. and {Sadowski}, G. and {Sagrist{\`a} Sell{\'e}s}, T. and {Sahlmann}, J. and {Salgado}, J. and {Salguero}, E. and {Sarasso}, M. and {Savietto}, H. and {Schnorhk}, A. and {Schultheis}, M. and {Sciacca}, E. and {Segol}, M. and {Segovia}, J.~C. and {Segransan}, D. and {Serpell}, E. and {Shih}, I. -C. and {Smareglia}, R. and {Smart}, R.~L. and {Smith}, C. and {Solano}, E. and {Solitro}, F. and {Sordo}, R. and {Soria Nieto}, S. and {Souchay}, J. and {Spagna}, A. and {Spoto}, F. and {Stampa}, U. and {Steele}, I.~A. and {Steidelm{\"u}ller}, H. and {Stephenson}, C.~A. and {Stoev}, H. and {Suess}, F.~F. and {S{\"u}veges}, M. and {Surdej}, J. and {Szabados}, L. and {Szegedi-Elek}, E. and {Tapiador}, D. and {Taris}, F. and {Tauran}, G. and {Taylor}, M.~B. and {Teixeira}, R. and {Terrett}, D. and {Tingley}, B. and {Trager}, S.~C. and {Turon}, C. and {Ulla}, A. and {Utrilla}, E. and {Valentini}, G. and {van Elteren}, A. and {Van Hemelryck}, E. and {van Leeuwen}, M. and {Varadi}, M. and {Vecchiato}, A. and {Veljanoski}, J. and {Via}, T. and {Vicente}, D. and {Vogt}, S. and {Voss}, H. and {Votruba}, V. and {Voutsinas}, S. and {Walmsley}, G. and {Weiler}, M. and {Weingrill}, K. and {Werner}, D. and {Wevers}, T. and {Whitehead}, G. and {Wyrzykowski}, {\L}. and {Yoldas}, A. and {{\v{Z}}erjal}, M. and {Zucker}, S. and {Zurbach}, C. and {Zwitter}, T. and {Alecu}, A. and {Allen}, M. and {Allende Prieto}, C. and {Amorim}, A. and {Anglada-Escud{\'e}}, G. and {Arsenijevic}, V. and {Azaz}, S. and {Balm}, P. and {Beck}, M. and {Bernstein}, H. -H. and {Bigot}, L. and {Bijaoui}, A. and {Blasco}, C. and {Bonfigli}, M. and {Bono}, G. and {Boudreault}, S. and {Bressan}, A. and {Brown}, S. and {Brunet}, P. -M. and {Bunclark}, P. and {Buonanno}, R. and {Butkevich}, A.~G. and {Carret}, C. and {Carrion}, C. and {Chemin}, L. and {Ch{\'e}reau}, F. and {Corcione}, L. and {Darmigny}, E. and {de Boer}, K.~S. and {de Teodoro}, P. and {de Zeeuw}, P.~T. and {Delle Luche}, C. and {Domingues}, C.~D. and {Dubath}, P. and {Fodor}, F. and {Fr{\'e}zouls}, B. and {Fries}, A. and {Fustes}, D. and {Fyfe}, D. and {Gallardo}, E. and {Gallegos}, J. and {Gardiol}, D. and {Gebran}, M. and {Gomboc}, A. and {G{\'o}mez}, A. and {Grux}, E. and {Gueguen}, A. and {Heyrovsky}, A. and {Hoar}, J. and {Iannicola}, G. and {Isasi Parache}, Y. and {Janotto}, A. -M. and {Joliet}, E. and {Jonckheere}, A. and {Keil}, R. and {Kim}, D. -W. and {Klagyivik}, P. and {Klar}, J. and {Knude}, J. and {Kochukhov}, O. and {Kolka}, I. and {Kos}, J. and {Kutka}, A. and {Lainey}, V. and {LeBouquin}, D. and {Liu}, C. and {Loreggia}, D. and {Makarov}, V.~V. and {Marseille}, M.~G. and {Martayan}, C. and {Martinez-Rubi}, O. and {Massart}, B. and {Meynadier}, F. and {Mignot}, S. and {Munari}, U. and {Nguyen}, A. -T. and {Nordlander}, T. and {Ocvirk}, P. and {O'Flaherty}, K.~S. and {Olias Sanz}, A. and {Ortiz}, P. and {Osorio}, J. and {Oszkiewicz}, D. and {Ouzounis}, A. and {Palmer}, M. and {Park}, P. and {Pasquato}, E. and {Peltzer}, C. and {Peralta}, J. and {P{\'e}turaud}, F. and {Pieniluoma}, T. and {Pigozzi}, E. and {Poels}, J. and {Prat}, G. and {Prod'homme}, T. and {Raison}, F. and {Rebordao}, J.~M. and {Risquez}, D. and {Rocca-Volmerange}, B. and {Rosen}, S. and {Ruiz-Fuertes}, M.~I. and {Russo}, F. and {Sembay}, S. and {Serraller Vizcaino}, I. and {Short}, A. and {Siebert}, A. and {Silva}, H. and {Sinachopoulos}, D. and {Slezak}, E. and {Soffel}, M. and {Sosnowska}, D. and {Strai{\v{z}}ys}, V. and {ter Linden}, M. and {Terrell}, D. and {Theil}, S. and {Tiede}, C. and {Troisi}, L. and {Tsalmantza}, P. and {Tur}, D. and {Vaccari}, M. and {Vachier}, F. and {Valles}, P. and {Van Hamme}, W. and {Veltz}, L. and {Virtanen}, J. and {Wallut}, J. -M. and {Wichmann}, R. and {Wilkinson}, M.~I. and {Ziaeepour}, H. and {Zschocke}, S.},
        title = "{The Gaia mission}",
      journal = {\aap},
         year = 2016,
        month = nov,
       volume = {595},
          eid = {A1},
        pages = {A1},
          doi = {10.1051/0004-6361/201629272},
archivePrefix = {arXiv},
       eprint = {1609.04153},
 primaryClass = {astro-ph.IM},
       adsurl = {https://ui.adsabs.harvard.edu/abs/2016A&A...595A...1G}
}

@ARTICLE{Spergel2015,
       author = {{Spergel}, D. and {Gehrels}, N. and {Baltay}, C. and {Bennett}, D. and {Breckinridge}, J. and {Donahue}, M. and {Dressler}, A. and {Gaudi}, B.~S. and {Greene}, T. and {Guyon}, O. and {Hirata}, C. and {Kalirai}, J. and {Kasdin}, N.~J. and {Macintosh}, B. and {Moos}, W. and {Perlmutter}, S. and {Postman}, M. and {Rauscher}, B. and {Rhodes}, J. and {Wang}, Y. and {Weinberg}, D. and {Benford}, D. and {Hudson}, M. and {Jeong}, W. -S. and {Mellier}, Y. and {Traub}, W. and {Yamada}, T. and {Capak}, P. and {Colbert}, J. and {Masters}, D. and {Penny}, M. and {Savransky}, D. and {Stern}, D. and {Zimmerman}, N. and {Barry}, R. and {Bartusek}, L. and {Carpenter}, K. and {Cheng}, E. and {Content}, D. and {Dekens}, F. and {Demers}, R. and {Grady}, K. and {Jackson}, C. and {Kuan}, G. and {Kruk}, J. and {Melton}, M. and {Nemati}, B. and {Parvin}, B. and {Poberezhskiy}, I. and {Peddie}, C. and {Ruffa}, J. and {Wallace}, J.~K. and {Whipple}, A. and {Wollack}, E. and {Zhao}, F.},
        title = "{Wide-Field InfrarRed Survey Telescope-Astrophysics Focused Telescope Assets WFIRST-AFTA 2015 Report}",
      journal = {arXiv e-prints},
         year = 2015,
        month = mar,
          eid = {arXiv:1503.03757},
        pages = {arXiv:1503.03757},
archivePrefix = {arXiv},
       eprint = {1503.03757},
 primaryClass = {astro-ph.IM},
       adsurl = {https://ui.adsabs.harvard.edu/abs/2015arXiv150303757S}
}

@misc{Foreman-Mackey2021,
  author = {Daniel Foreman-Mackey and Arjun Savel and Rodrigo Luger and
            Ian Czekala and Eric Agol and Adrian Price-Whelan and
            Christina Hedges and Emily Gilbert and Tom Barclay and Luke Bouma
            and Timothy D. Brandt},
   title = {exoplanet-dev/exoplanet v0.4.5},
   month = mar,
    year = 2021,
     doi = {10.5281/zenodo.1998447},
     url = {https://doi.org/10.5281/zenodo.1998447}
}

@article{Salvatier2016,
    title={Probabilistic programming in Python using PyMC3},
   author={Salvatier, John and Wiecki, Thomas V and Fonnesbeck, Christopher},
  journal={PeerJ Computer Science},
   volume={2},
    pages={e55},
     year={2016},
publisher={PeerJ Inc.}
}

@ARTICLE{Kipping2010,
       author = {{Kipping}, David M. and {Tinetti}, Giovanna},
        title = "{Nightside pollution of exoplanet transit depths}",
      journal = {\mnras},
         year = 2010,
        month = oct,
       volume = {407},
       number = {4},
        pages = {2589-2598},
          doi = {10.1111/j.1365-2966.2010.17094.x},
archivePrefix = {arXiv},
       eprint = {0912.1133},
 primaryClass = {astro-ph.EP},
       adsurl = {https://ui.adsabs.harvard.edu/abs/2010MNRAS.407.2589K}
}

@article{numpy,
 author = {Walt, Stefan van der and Colbert, S. Chris and Varoquaux, Gael},
 title = {The NumPy Array: A Structure for Efficient Numerical Computation},
 journal = {Computing in Science and Eng.},
 issue_date = {March 2011},
 volume = {13},
 number = {2},
 month = mar,
 year = {2011},
 issn = {1521-9615},
 pages = {22--30},
 numpages = {9},
 url = {https://doi.org/10.1109/MCSE.2011.37},
 doi = {10.1109/MCSE.2011.37},
 acmid = {1957466},
 publisher = {IEEE Educational Activities Department},
 address = {Piscataway, NJ, USA}
}

@Misc{scipy,
  author =    {Eric Jones and Travis Oliphant and Pearu Peterson and others},
  title =     {{SciPy}: Open source scientific tools for {Python}},
  year =      {2001},
  url = "http://www.scipy.org/",
  note = {[Online; accessed <today>]}
}

@ARTICLE{matplotlib,
   author = {{Hunter}, J.~D.},
    title = "{Matplotlib: A 2D Graphics Environment}",
  journal = {Computing in Science and Engineering},
     year = 2007,
    month = may,
   volume = 9,
    pages = {90-95},
      doi = {10.1109/MCSE.2007.55},
   adsurl = {http://adsabs.harvard.edu/abs/2007CSE.....9...90H}
}

@ARTICLE{Borucki2010,
       author = {{Borucki}, William J. and {Koch}, David and {Basri}, Gibor and {Batalha}, Natalie and {Brown}, Timothy and {Caldwell}, Douglas and {Caldwell}, John and {Christensen-Dalsgaard}, J{\o}rgen and {Cochran}, William D. and {DeVore}, Edna and {Dunham}, Edward W. and {Dupree}, Andrea K. and {Gautier}, Thomas N. and {Geary}, John C. and {Gilliland}, Ronald and {Gould}, Alan and {Howell}, Steve B. and {Jenkins}, Jon M. and {Kondo}, Yoji and {Latham}, David W. and {Marcy}, Geoffrey W. and {Meibom}, S{\o}ren and {Kjeldsen}, Hans and {Lissauer}, Jack J. and {Monet}, David G. and {Morrison}, David and {Sasselov}, Dimitar and {Tarter}, Jill and {Boss}, Alan and {Brownlee}, Don and {Owen}, Toby and {Buzasi}, Derek and {Charbonneau}, David and {Doyle}, Laurance and {Fortney}, Jonathan and {Ford}, Eric B. and {Holman}, Matthew J. and {Seager}, Sara and {Steffen}, Jason H. and {Welsh}, William F. and {Rowe}, Jason and {Anderson}, Howard and {Buchhave}, Lars and {Ciardi}, David and {Walkowicz}, Lucianne and {Sherry}, William and {Horch}, Elliott and {Isaacson}, Howard and {Everett}, Mark E. and {Fischer}, Debra and {Torres}, Guillermo and {Johnson}, John Asher and {Endl}, Michael and {MacQueen}, Phillip and {Bryson}, Stephen T. and {Dotson}, Jessie and {Haas}, Michael and {Kolodziejczak}, Jeffrey and {Van Cleve}, Jeffrey and {Chandrasekaran}, Hema and {Twicken}, Joseph D. and {Quintana}, Elisa V. and {Clarke}, Bruce D. and {Allen}, Christopher and {Li}, Jie and {Wu}, Haley and {Tenenbaum}, Peter and {Verner}, Ekaterina and {Bruhweiler}, Frederick and {Barnes}, Jason and {Prsa}, Andrej},
        title = "{Kepler Planet-Detection Mission: Introduction and First Results}",
      journal = {Science},
         year = 2010,
        month = feb,
       volume = {327},
       number = {5968},
        pages = {977},
          doi = {10.1126/science.1185402},
       adsurl = {https://ui.adsabs.harvard.edu/abs/2010Sci...327..977B}
}

@ARTICLE{Ricker2015,
       author = {{Ricker}, George R. and {Winn}, Joshua N. and {Vanderspek}, Roland and {Latham}, David W. and {Bakos}, G{\'a}sp{\'a}r {\'A}. and {Bean}, Jacob L. and {Berta-Thompson}, Zachory K. and {Brown}, Timothy M. and {Buchhave}, Lars and {Butler}, Nathaniel R. and {Butler}, R. Paul and {Chaplin}, William J. and {Charbonneau}, David and {Christensen-Dalsgaard}, J{\o}rgen and {Clampin}, Mark and {Deming}, Drake and {Doty}, John and {De Lee}, Nathan and {Dressing}, Courtney and {Dunham}, Edward W. and {Endl}, Michael and {Fressin}, Francois and {Ge}, Jian and {Henning}, Thomas and {Holman}, Matthew J. and {Howard}, Andrew W. and {Ida}, Shigeru and {Jenkins}, Jon M. and {Jernigan}, Garrett and {Johnson}, John Asher and {Kaltenegger}, Lisa and {Kawai}, Nobuyuki and {Kjeldsen}, Hans and {Laughlin}, Gregory and {Levine}, Alan M. and {Lin}, Douglas and {Lissauer}, Jack J. and {MacQueen}, Phillip and {Marcy}, Geoffrey and {McCullough}, Peter R. and {Morton}, Timothy D. and {Narita}, Norio and {Paegert}, Martin and {Palle}, Enric and {Pepe}, Francesco and {Pepper}, Joshua and {Quirrenbach}, Andreas and {Rinehart}, Stephen A. and {Sasselov}, Dimitar and {Sato}, Bun'ei and {Seager}, Sara and {Sozzetti}, Alessandro and {Stassun}, Keivan G. and {Sullivan}, Peter and {Szentgyorgyi}, Andrew and {Torres}, Guillermo and {Udry}, Stephane and {Villasenor}, Joel},
        title = "{Transiting Exoplanet Survey Satellite (TESS)}",
      journal = {Journal of Astronomical Telescopes, Instruments, and Systems},
         year = 2015,
        month = jan,
       volume = {1},
          eid = {014003},
        pages = {014003},
          doi = {10.1117/1.JATIS.1.1.014003},
       adsurl = {https://ui.adsabs.harvard.edu/abs/2015JATIS...1a4003R}
}

@ARTICLE{Foreman-Mackey2016,
       author = {{Foreman-Mackey}, Daniel},
        title = "{corner.py: Scatterplot matrices in Python}",
      journal = {The Journal of Open Source Software},
         year = 2016,
        month = jun,
       volume = {1},
        pages = {24},
          doi = {10.21105/joss.00024},
       adsurl = {https://ui.adsabs.harvard.edu/abs/2016JOSS....1...24F}
}

@article{exoplanet:joss,
       author = {{Foreman-Mackey}, Daniel and {Luger}, Rodrigo and {Agol}, Eric
                and {Barclay}, Thomas and {Bouma}, Luke G. and {Brandt},
                Timothy D. and {Czekala}, Ian and {David}, Trevor J. and
                {Dong}, Jiayin and {Gilbert}, Emily A. and {Gordon}, Tyler A.
                and {Hedges}, Christina and {Hey}, Daniel R. and {Morris},
                Brett M. and {Price-Whelan}, Adrian M. and {Savel}, Arjun B.},
        title = "{exoplanet: Gradient-based probabilistic inference for
                  exoplanet data \\& other astronomical time series}",
      journal = {arXiv e-prints},
         year = 2021,
        month = may,
          eid = {arXiv:2105.01994},
        pages = {arXiv:2105.01994},
        archivePrefix = {arXiv},
       eprint = {2105.01994},
 primaryClass = {astro-ph.IM},
       adsurl = {https://ui.adsabs.harvard.edu/abs/2021arXiv210501994F}
      }

@misc{exoplanet:zenodo,
  author = {Daniel Foreman-Mackey and Arjun Savel and Rodrigo Luger  and
            Eric Agol and Ian Czekala and Adrian Price-Whelan and
            Christina Hedges and Emily Gilbert and Luke Bouma
            and Timothy D. Brandt and Tom Barclay},
   title = {exoplanet-dev/exoplanet v0.5.1},
   month = jun,
    year = 2021,
     doi = {10.5281/zenodo.1998447},
     url = {https://doi.org/10.5281/zenodo.1998447}
     }

@article{exoplanet:pymc3,
    title={Probabilistic programming in Python using PyMC3},
   author={Salvatier, John and Wiecki, Thomas V and Fonnesbeck, Christopher},
  journal={PeerJ Computer Science},
   volume={2},
    pages={e55},
     year={2016},
     publisher={PeerJ Inc.}
}

@article{exoplanet:theano,
    title="{Theano: A {Python} framework for fast computation of mathematical
            expressions}",
   author={{Theano Development Team}},
  journal={arXiv e-prints},
   volume={abs/1605.02688},
     year=2016,
    month=may,
      url={http://arxiv.org/abs/1605.02688}
      }

@article{exoplanet:arviz,
    title={{ArviZ} a unified library for exploratory analysis of {Bayesian}
           models in {Python}},
   author={Kumar, Ravin and Carroll, Colin and Hartikainen, Ari and Martin,
           Osvaldo A.},
  journal={The Journal of Open Source Software},
     year=2019,
      doi={10.21105/joss.01143},
      url={http://joss.theoj.org/papers/10.21105/joss.01143}
      }

@ARTICLE{exoplanet:kipping13b,
       author = {{Kipping}, D.~M.},
        title = "{Parametrizing the exoplanet eccentricity distribution with
                  the beta  distribution.}",
      journal = {\\mnras},
         year = "2013",
        month = jul,
       volume = 434,
        pages = {L51-L55},
          doi = {10.1093/mnrasl/slt075},
       adsurl = {https://ui.adsabs.harvard.edu/abs/2013MNRAS.434L..51K}
      }

@article{exoplanet:astropy13,
   author = {{Astropy Collaboration} and {Robitaille}, T.~P. and {Tollerud},
             E.~J. and {Greenfield}, P. and {Droettboom}, M. and {Bray}, E. and
             {Aldcroft}, T. and {Davis}, M. and {Ginsburg}, A. and
             {Price-Whelan}, A.~M. and {Kerzendorf}, W.~E. and {Conley}, A. and
             {Crighton}, N. and {Barbary}, K. and {Muna}, D. and {Ferguson}, H.
             and {Grollier}, F. and {Parikh}, M.~M. and {Nair}, P.~H. and
             {Unther}, H.~M. and {Deil}, C. and {Woillez}, J. and {Conseil}, S.
             and {Kramer}, R. and {Turner}, J.~E.~H. and {Singer}, L. and
             {Fox}, R. and {Weaver}, B.~A. and {Zabalza}, V. and {Edwards},
             Z.~I. and {Azalee Bostroem}, K. and {Burke}, D.~J. and {Casey},
             A.~R. and {Crawford}, S.~M. and {Dencheva}, N. and {Ely}, J. and
             {Jenness}, T. and {Labrie}, K. and {Lim}, P.~L. and
             {Pierfederici}, F. and {Pontzen}, A. and {Ptak}, A. and {Refsdal},
             B. and {Servillat}, M. and {Streicher}, O.},
    title = "{Astropy: A community Python package for astronomy}",
  journal = {\\aap},
     year = 2013,
    month = oct,
   volume = 558,
    pages = {A33},
      doi = {10.1051/0004-6361/201322068},
   adsurl = {http://adsabs.harvard.edu/abs/2013A%26A...558A..33A}
  }

@article{exoplanet:astropy18,
   author = {{Astropy Collaboration} and {Price-Whelan}, A.~M. and
             {Sip{\\H o}cz}, B.~M. and {G{\\"u}nther}, H.~M. and {Lim}, P.~L. and
             {Crawford}, S.~M. and {Conseil}, S. and {Shupe}, D.~L. and
             {Craig}, M.~W. and {Dencheva}, N. and {Ginsburg}, A. and
             {VanderPlas}, J.~T. and {Bradley}, L.~D. and
             {P{\\\'e}rez-Su{\\\'a}rez}, D. and {de Val-Borro}, M.
             and {Aldcroft}, T.~L. and {Cruz}, K.~L. and {Robitaille}, T.~P.
             and {Tollerud}, E.~J. and {Ardelean}, C. and {Babej}, T. and
             {Bach}, Y.~P. and {Bachetti}, M. and {Bakanov}, A.~V. and
             {Bamford}, S.~P. and {Barentsen}, G. and {Barmby}, P. and
             {Baumbach}, A. and {Berry}, K.~L.  and {Biscani}, F. and
             {Boquien}, M. and {Bostroem}, K.~A. and {Bouma}, L.~G. and
             {Brammer}, G.~B. and {Bray}, E.~M. and {Breytenbach}, H. and
             {Buddelmeijer}, H. and {Burke}, D.~J. and {Calderone}, G. and
             {Cano Rodr{\\\'{\\i}}guez}, J.~L. and {Cara}, M. and {Cardoso},
             J.~V.~M. and {Cheedella}, S. and {Copin}, Y. and {Corrales}, L.
             and {Crichton}, D. and {D\'Avella}, D. and {Deil}, C. and
             {Depagne}, {\\\'E}. and {Dietrich}, J.~P. and {Donath}, A. and
             {Droettboom}, M. and {Earl}, N. and {Erben}, T. and {Fabbro}, S.
             and {Ferreira}, L.~A. and {Finethy}, T. and {Fox}, R.~T. and
             {Garrison}, L.~H. and {Gibbons}, S.~L.~J. and {Goldstein}, D.~A.
             and {Gommers}, R. and {Greco}, J.~P. and {Greenfield}, P. and
             {Groener}, A.~M. and {Grollier}, F. and {Hagen}, A. and {Hirst},
             P. and {Homeier}, D. and {Horton}, A.~J. and {Hosseinzadeh}, G.
             and {Hu}, L. and {Hunkeler}, J.~S. and {Ivezi{\\\'c}}, {\\v Z}. and
             {Jain}, A. and {Jenness}, T. and {Kanarek}, G. and {Kendrew}, S.
             and {Kern}, N.~S. and {Kerzendorf}, W.~E. and {Khvalko}, A. and
             {King}, J. and {Kirkby}, D. and {Kulkarni}, A.~M. and {Kumar}, A.
             and {Lee}, A.  and {Lenz}, D.  and {Littlefair}, S.~P. and {Ma},
             Z. and {Macleod}, D.~M. and {Mastropietro}, M. and {McCully}, C.
             and {Montagnac}, S. and {Morris}, B.~M. and {Mueller}, M. and
             {Mumford}, S.~J. and {Muna}, D. and {Murphy}, N.~A. and {Nelson},
             S. and {Nguyen}, G.~H. and {Ninan}, J.~P. and {N{\\"o}the}, M. and
             {Ogaz}, S. and {Oh}, S. and {Parejko}, J.~K.  and {Parley}, N. and
             {Pascual}, S. and {Patil}, R. and {Patil}, A.~A.  and {Plunkett},
             A.~L. and {Prochaska}, J.~X. and {Rastogi}, T. and {Reddy Janga},
             V. and {Sabater}, J.  and {Sakurikar}, P. and {Seifert}, M. and
             {Sherbert}, L.~E. and {Sherwood-Taylor}, H. and {Shih}, A.~Y. and
             {Sick}, J. and {Silbiger}, M.~T. and {Singanamalla}, S. and
             {Singer}, L.~P. and {Sladen}, P.~H. and {Sooley}, K.~A. and
             {Sornarajah}, S. and {Streicher}, O. and {Teuben}, P. and
             {Thomas}, S.~W. and {Tremblay}, G.~R. and {Turner}, J.~E.~H. and
             {Terr{\\\'o}n}, V.  and {van Kerkwijk}, M.~H. and {de la Vega}, A.
             and {Watkins}, L.~L. and {Weaver}, B.~A. and {Whitmore}, J.~B. and
             {Woillez}, J.  and {Zabalza}, V. and {Astropy Contributors}},
    title = "{The Astropy Project: Building an Open-science Project and Status
              of the v2.0 Core Package}",
  journal = {\\aj},
     year = 2018,
    month = sep,
   volume = 156,
    pages = {123},
      doi = {10.3847/1538-3881/aabc4f},
   adsurl = {http://adsabs.harvard.edu/abs/2018AJ....156..123A}
  }

@ARTICLE{Bakos2004,
       author = {{Bakos}, G. and {Noyes}, R.~W. and {Kov{\'a}cs}, G. and {Stanek}, K.~Z. and {Sasselov}, D.~D. and {Domsa}, I.},
        title = "{Wide-Field Millimagnitude Photometry with the HAT: A Tool for Extrasolar Planet Detection}",
      journal = {\pasp},
         year = 2004,
        month = mar,
       volume = {116},
       number = {817},
        pages = {266-277},
          doi = {10.1086/382735},
archivePrefix = {arXiv},
       eprint = {astro-ph/0401219},
 primaryClass = {astro-ph},
       adsurl = {https://ui.adsabs.harvard.edu/abs/2004PASP..116..266B}
}

@ARTICLE{Pollacco2006,
       author = {{Pollacco}, D.~L. and {Skillen}, I. and {Collier Cameron}, A. and {Christian}, D.~J. and {Hellier}, C. and {Irwin}, J. and {Lister}, T.~A. and {Street}, R.~A. and {West}, R.~G. and {Anderson}, D.~R. and {Clarkson}, W.~I. and {Deeg}, H. and {Enoch}, B. and {Evans}, A. and {Fitzsimmons}, A. and {Haswell}, C.~A. and {Hodgkin}, S. and {Horne}, K. and {Kane}, S.~R. and {Keenan}, F.~P. and {Maxted}, P.~F.~L. and {Norton}, A.~J. and {Osborne}, J. and {Parley}, N.~R. and {Ryans}, R.~S.~I. and {Smalley}, B. and {Wheatley}, P.~J. and {Wilson}, D.~M.},
        title = "{The WASP Project and the SuperWASP Cameras}",
      journal = {\pasp},
         year = 2006,
        month = oct,
       volume = {118},
       number = {848},
        pages = {1407-1418},
          doi = {10.1086/508556},
archivePrefix = {arXiv},
       eprint = {astro-ph/0608454},
 primaryClass = {astro-ph},
       adsurl = {https://ui.adsabs.harvard.edu/abs/2006PASP..118.1407P}
}

@ARTICLE{Pepper2007,
       author = {{Pepper}, Joshua and {Pogge}, Richard W. and {DePoy}, D.~L. and {Marshall}, J.~L. and {Stanek}, K.~Z. and {Stutz}, Amelia M. and {Poindexter}, Shawn and {Siverd}, Robert and {O'Brien}, Thomas P. and {Trueblood}, Mark and {Trueblood}, Patricia},
        title = "{The Kilodegree Extremely Little Telescope (KELT): A Small Robotic Telescope for Large-Area Synoptic Surveys}",
      journal = {\pasp},
         year = 2007,
        month = aug,
       volume = {119},
       number = {858},
        pages = {923-935},
          doi = {10.1086/521836},
archivePrefix = {arXiv},
       eprint = {0704.0460},
 primaryClass = {astro-ph},
       adsurl = {https://ui.adsabs.harvard.edu/abs/2007PASP..119..923P}
}

@ARTICLE{Ivezic2019,
       author = {{Ivezi{\'c}}, {\v{Z}}eljko and {Kahn}, Steven M. and {Tyson}, J. Anthony and {Abel}, Bob and {Acosta}, Emily and {Allsman}, Robyn and {Alonso}, David and {AlSayyad}, Yusra and {Anderson}, Scott F. and {Andrew}, John and {Angel}, James Roger P. and {Angeli}, George Z. and {Ansari}, Reza and {Antilogus}, Pierre and {Araujo}, Constanza and {Armstrong}, Robert and {Arndt}, Kirk T. and {Astier}, Pierre and {Aubourg}, {\'E}ric and {Auza}, Nicole and {Axelrod}, Tim S. and {Bard}, Deborah J. and {Barr}, Jeff D. and {Barrau}, Aurelian and {Bartlett}, James G. and {Bauer}, Amanda E. and {Bauman}, Brian J. and {Baumont}, Sylvain and {Bechtol}, Ellen and {Bechtol}, Keith and {Becker}, Andrew C. and {Becla}, Jacek and {Beldica}, Cristina and {Bellavia}, Steve and {Bianco}, Federica B. and {Biswas}, Rahul and {Blanc}, Guillaume and {Blazek}, Jonathan and {Blandford}, Roger D. and {Bloom}, Josh S. and {Bogart}, Joanne and {Bond}, Tim W. and {Booth}, Michael T. and {Borgland}, Anders W. and {Borne}, Kirk and {Bosch}, James F. and {Boutigny}, Dominique and {Brackett}, Craig A. and {Bradshaw}, Andrew and {Brandt}, William Nielsen and {Brown}, Michael E. and {Bullock}, James S. and {Burchat}, Patricia and {Burke}, David L. and {Cagnoli}, Gianpietro and {Calabrese}, Daniel and {Callahan}, Shawn and {Callen}, Alice L. and {Carlin}, Jeffrey L. and {Carlson}, Erin L. and {Chandrasekharan}, Srinivasan and {Charles-Emerson}, Glenaver and {Chesley}, Steve and {Cheu}, Elliott C. and {Chiang}, Hsin-Fang and {Chiang}, James and {Chirino}, Carol and {Chow}, Derek and {Ciardi}, David R. and {Claver}, Charles F. and {Cohen-Tanugi}, Johann and {Cockrum}, Joseph J. and {Coles}, Rebecca and {Connolly}, Andrew J. and {Cook}, Kem H. and {Cooray}, Asantha and {Covey}, Kevin R. and {Cribbs}, Chris and {Cui}, Wei and {Cutri}, Roc and {Daly}, Philip N. and {Daniel}, Scott F. and {Daruich}, Felipe and {Daubard}, Guillaume and {Daues}, Greg and {Dawson}, William and {Delgado}, Francisco and {Dellapenna}, Alfred and {de Peyster}, Robert and {de Val-Borro}, Miguel and {Digel}, Seth W. and {Doherty}, Peter and {Dubois}, Richard and {Dubois-Felsmann}, Gregory P. and {Durech}, Josef and {Economou}, Frossie and {Eifler}, Tim and {Eracleous}, Michael and {Emmons}, Benjamin L. and {Fausti Neto}, Angelo and {Ferguson}, Henry and {Figueroa}, Enrique and {Fisher-Levine}, Merlin and {Focke}, Warren and {Foss}, Michael D. and {Frank}, James and {Freemon}, Michael D. and {Gangler}, Emmanuel and {Gawiser}, Eric and {Geary}, John C. and {Gee}, Perry and {Geha}, Marla and {Gessner}, Charles J.~B. and {Gibson}, Robert R. and {Gilmore}, D. Kirk and {Glanzman}, Thomas and {Glick}, William and {Goldina}, Tatiana and {Goldstein}, Daniel A. and {Goodenow}, Iain and {Graham}, Melissa L. and {Gressler}, William J. and {Gris}, Philippe and {Guy}, Leanne P. and {Guyonnet}, Augustin and {Haller}, Gunther and {Harris}, Ron and {Hascall}, Patrick A. and {Haupt}, Justine and {Hernandez}, Fabio and {Herrmann}, Sven and {Hileman}, Edward and {Hoblitt}, Joshua and {Hodgson}, John A. and {Hogan}, Craig and {Howard}, James D. and {Huang}, Dajun and {Huffer}, Michael E. and {Ingraham}, Patrick and {Innes}, Walter R. and {Jacoby}, Suzanne H. and {Jain}, Bhuvnesh and {Jammes}, Fabrice and {Jee}, M. James and {Jenness}, Tim and {Jernigan}, Garrett and {Jevremovi{\'c}}, Darko and {Johns}, Kenneth and {Johnson}, Anthony S. and {Johnson}, Margaret W.~G. and {Jones}, R. Lynne and {Juramy-Gilles}, Claire and {Juri{\'c}}, Mario and {Kalirai}, Jason S. and {Kallivayalil}, Nitya J. and {Kalmbach}, Bryce and {Kantor}, Jeffrey P. and {Karst}, Pierre and {Kasliwal}, Mansi M. and {Kelly}, Heather and {Kessler}, Richard and {Kinnison}, Veronica and {Kirkby}, David and {Knox}, Lloyd and {Kotov}, Ivan V. and {Krabbendam}, Victor L. and {Krughoff}, K. Simon and {Kub{\'a}nek}, Petr and {Kuczewski}, John and {Kulkarni}, Shri and {Ku}, John and {Kurita}, Nadine R. and {Lage}, Craig S. and {Lambert}, Ron and {Lange}, Travis and {Langton}, J. Brian and {Le Guillou}, Laurent and {Levine}, Deborah and {Liang}, Ming and {Lim}, Kian-Tat and {Lintott}, Chris J. and {Long}, Kevin E. and {Lopez}, Margaux and {Lotz}, Paul J. and {Lupton}, Robert H. and {Lust}, Nate B. and {MacArthur}, Lauren A. and {Mahabal}, Ashish and {Mandelbaum}, Rachel and {Markiewicz}, Thomas W. and {Marsh}, Darren S. and {Marshall}, Philip J. and {Marshall}, Stuart and {May}, Morgan and {McKercher}, Robert and {McQueen}, Michelle and {Meyers}, Joshua and {Migliore}, Myriam and {Miller}, Michelle and {Mills}, David J. and {Miraval}, Connor and {Moeyens}, Joachim and {Moolekamp}, Fred E. and {Monet}, David G. and {Moniez}, Marc and {Monkewitz}, Serge and {Montgomery}, Christopher and {Morrison}, Christopher B. and {Mueller}, Fritz and {Muller}, Gary P. and {Mu{\~n}oz Arancibia}, Freddy and {Neill}, Douglas R. and {Newbry}, Scott P. and {Nief}, Jean-Yves and {Nomerotski}, Andrei and {Nordby}, Martin and {O'Connor}, Paul and {Oliver}, John and {Olivier}, Scot S. and {Olsen}, Knut and {O'Mullane}, William and {Ortiz}, Sandra and {Osier}, Shawn and {Owen}, Russell E. and {Pain}, Reynald and {Palecek}, Paul E. and {Parejko}, John K. and {Parsons}, James B. and {Pease}, Nathan M. and {Peterson}, J. Matt and {Peterson}, John R. and {Petravick}, Donald L. and {Libby Petrick}, M.~E. and {Petry}, Cathy E. and {Pierfederici}, Francesco and {Pietrowicz}, Stephen and {Pike}, Rob and {Pinto}, Philip A. and {Plante}, Raymond and {Plate}, Stephen and {Plutchak}, Joel P. and {Price}, Paul A. and {Prouza}, Michael and {Radeka}, Veljko and {Rajagopal}, Jayadev and {Rasmussen}, Andrew P. and {Regnault}, Nicolas and {Reil}, Kevin A. and {Reiss}, David J. and {Reuter}, Michael A. and {Ridgway}, Stephen T. and {Riot}, Vincent J. and {Ritz}, Steve and {Robinson}, Sean and {Roby}, William and {Roodman}, Aaron and {Rosing}, Wayne and {Roucelle}, Cecille and {Rumore}, Matthew R. and {Russo}, Stefano and {Saha}, Abhijit and {Sassolas}, Benoit and {Schalk}, Terry L. and {Schellart}, Pim and {Schindler}, Rafe H. and {Schmidt}, Samuel and {Schneider}, Donald P. and {Schneider}, Michael D. and {Schoening}, William and {Schumacher}, German and {Schwamb}, Megan E. and {Sebag}, Jacques and {Selvy}, Brian and {Sembroski}, Glenn H. and {Seppala}, Lynn G. and {Serio}, Andrew and {Serrano}, Eduardo and {Shaw}, Richard A. and {Shipsey}, Ian and {Sick}, Jonathan and {Silvestri}, Nicole and {Slater}, Colin T. and {Smith}, J. Allyn and {Smith}, R. Chris and {Sobhani}, Shahram and {Soldahl}, Christine and {Storrie-Lombardi}, Lisa and {Stover}, Edward and {Strauss}, Michael A. and {Street}, Rachel A. and {Stubbs}, Christopher W. and {Sullivan}, Ian S. and {Sweeney}, Donald and {Swinbank}, John D. and {Szalay}, Alexander and {Takacs}, Peter and {Tether}, Stephen A. and {Thaler}, Jon J. and {Thayer}, John Gregg and {Thomas}, Sandrine and {Thornton}, Adam J. and {Thukral}, Vaikunth and {Tice}, Jeffrey and {Trilling}, David E. and {Turri}, Max and {Van Berg}, Richard and {Vanden Berk}, Daniel and {Vetter}, Kurt and {Virieux}, Francoise and {Vucina}, Tomislav and {Wahl}, William and {Walkowicz}, Lucianne and {Walsh}, Brian and {Walter}, Christopher W. and {Wang}, Daniel L. and {Wang}, Shin-Yawn and {Warner}, Michael and {Wiecha}, Oliver and {Willman}, Beth and {Winters}, Scott E. and {Wittman}, David and {Wolff}, Sidney C. and {Wood-Vasey}, W. Michael and {Wu}, Xiuqin and {Xin}, Bo and {Yoachim}, Peter and {Zhan}, Hu},
        title = "{LSST: From Science Drivers to Reference Design and Anticipated Data Products}",
      journal = {\apj},
         year = 2019,
        month = mar,
       volume = {873},
       number = {2},
          eid = {111},
        pages = {111},
          doi = {10.3847/1538-4357/ab042c},
archivePrefix = {arXiv},
       eprint = {0805.2366},
 primaryClass = {astro-ph},
       adsurl = {https://ui.adsabs.harvard.edu/abs/2019ApJ...873..111I}
}

@ARTICLE{Auvergne2009,
       author = {{Auvergne}, M. and {Bodin}, P. and {Boisnard}, L. and {Buey}, J. -T. and {Chaintreuil}, S. and {Epstein}, G. and {Jouret}, M. and {Lam-Trong}, T. and {Levacher}, P. and {Magnan}, A. and {Perez}, R. and {Plasson}, P. and {Plesseria}, J. and {Peter}, G. and {Steller}, M. and {Tiph{\`e}ne}, D. and {Baglin}, A. and {Agogu{\'e}}, P. and {Appourchaux}, T. and {Barbet}, D. and {Beaufort}, T. and {Bellenger}, R. and {Berlin}, R. and {Bernardi}, P. and {Blouin}, D. and {Boumier}, P. and {Bonneau}, F. and {Briet}, R. and {Butler}, B. and {Cautain}, R. and {Chiavassa}, F. and {Costes}, V. and {Cuvilho}, J. and {Cunha-Parro}, V. and {de Oliveira Fialho}, F. and {Decaudin}, M. and {Defise}, J. -M. and {Djalal}, S. and {Docclo}, A. and {Drummond}, R. and {Dupuis}, O. and {Exil}, G. and {Faur{\'e}}, C. and {Gaboriaud}, A. and {Gamet}, P. and {Gavalda}, P. and {Grolleau}, E. and {Gueguen}, L. and {Guivarc'h}, V. and {Guterman}, P. and {Hasiba}, J. and {Huntzinger}, G. and {Hustaix}, H. and {Imbert}, C. and {Jeanville}, G. and {Johlander}, B. and {Jorda}, L. and {Journoud}, P. and {Karioty}, F. and {Kerjean}, L. and {Lafond}, L. and {Lapeyrere}, V. and {Landiech}, P. and {Larqu{\'e}}, T. and {Laudet}, P. and {Le Merrer}, J. and {Leporati}, L. and {Leruyet}, B. and {Levieuge}, B. and {Llebaria}, A. and {Martin}, L. and {Mazy}, E. and {Mesnager}, J. -M. and {Michel}, J. -P. and {Moalic}, J. -P. and {Monjoin}, W. and {Naudet}, D. and {Neukirchner}, S. and {Nguyen-Kim}, K. and {Ollivier}, M. and {Orcesi}, J. -L. and {Ottacher}, H. and {Oulali}, A. and {Parisot}, J. and {Perruchot}, S. and {Piacentino}, A. and {Pinheiro da Silva}, L. and {Platzer}, J. and {Pontet}, B. and {Pradines}, A. and {Quentin}, C. and {Rohbeck}, U. and {Rolland}, G. and {Rollenhagen}, F. and {Romagnan}, R. and {Russ}, N. and {Samadi}, R. and {Schmidt}, R. and {Schwartz}, N. and {Sebbag}, I. and {Smit}, H. and {Sunter}, W. and {Tello}, M. and {Toulouse}, P. and {Ulmer}, B. and {Vandermarcq}, O. and {Vergnault}, E. and {Wallner}, R. and {Waultier}, G. and {Zanatta}, P.},
        title = "{The CoRoT satellite in flight: description and performance}",
      journal = {\aap},
         year = 2009,
        month = oct,
       volume = {506},
       number = {1},
        pages = {411-424},
          doi = {10.1051/0004-6361/200810860},
archivePrefix = {arXiv},
       eprint = {0901.2206},
 primaryClass = {astro-ph.SR},
       adsurl = {https://ui.adsabs.harvard.edu/abs/2009A&A...506..411A}
}

@ARTICLE{Howell2014,
       author = {{Howell}, Steve B. and {Sobeck}, Charlie and {Haas}, Michael and {Still}, Martin and {Barclay}, Thomas and {Mullally}, Fergal and {Troeltzsch}, John and {Aigrain}, Suzanne and {Bryson}, Stephen T. and {Caldwell}, Doug and {Chaplin}, William J. and {Cochran}, William D. and {Huber}, Daniel and {Marcy}, Geoffrey W. and {Miglio}, Andrea and {Najita}, Joan R. and {Smith}, Marcie and {Twicken}, J.~D. and {Fortney}, Jonathan J.},
        title = "{The K2 Mission: Characterization and Early Results}",
      journal = {\pasp},
         year = 2014,
        month = apr,
       volume = {126},
       number = {938},
        pages = {398},
          doi = {10.1086/676406},
archivePrefix = {arXiv},
       eprint = {1402.5163},
 primaryClass = {astro-ph.IM},
       adsurl = {https://ui.adsabs.harvard.edu/abs/2014PASP..126..398H}
}

@ARTICLE{Benz2021,
       author = {{Benz}, W. and {Broeg}, C. and {Fortier}, A. and {Rando}, N. and {Beck}, T. and {Beck}, M. and {Queloz}, D. and {Ehrenreich}, D. and {Maxted}, P.~F.~L. and {Isaak}, K.~G. and {Billot}, N. and {Alibert}, Y. and {Alonso}, R. and {Ant{\'o}nio}, C. and {Asquier}, J. and {Bandy}, T. and {B{\'a}rczy}, T. and {Barrado}, D. and {Barros}, S.~C.~C. and {Baumjohann}, W. and {Bekkelien}, A. and {Bergomi}, M. and {Biondi}, F. and {Bonfils}, X. and {Borsato}, L. and {Brandeker}, A. and {Busch}, M. -D. and {Cabrera}, J. and {Cessa}, V. and {Charnoz}, S. and {Chazelas}, B. and {Collier Cameron}, A. and {Corral Van Damme}, C. and {Cortes}, D. and {Davies}, M.~B. and {Deleuil}, M. and {Deline}, A. and {Delrez}, L. and {Demangeon}, O. and {Demory}, B.~O. and {Erikson}, A. and {Farinato}, J. and {Fossati}, L. and {Fridlund}, M. and {Futyan}, D. and {Gandolfi}, D. and {Garcia Munoz}, A. and {Gillon}, M. and {Guterman}, P. and {Gutierrez}, A. and {Hasiba}, J. and {Heng}, K. and {Hernandez}, E. and {Hoyer}, S. and {Kiss}, L.~L. and {Kovacs}, Z. and {Kuntzer}, T. and {Laskar}, J. and {Lecavelier des Etangs}, A. and {Lendl}, M. and {L{\'o}pez}, A. and {Lora}, I. and {Lovis}, C. and {L{\"u}ftinger}, T. and {Magrin}, D. and {Malvasio}, L. and {Marafatto}, L. and {Michaelis}, H. and {de Miguel}, D. and {Modrego}, D. and {Munari}, M. and {Nascimbeni}, V. and {Olofsson}, G. and {Ottacher}, H. and {Ottensamer}, R. and {Pagano}, I. and {Palacios}, R. and {Pall{\'e}}, E. and {Peter}, G. and {Piazza}, D. and {Piotto}, G. and {Pizarro}, A. and {Pollaco}, D. and {Ragazzoni}, R. and {Ratti}, F. and {Rauer}, H. and {Ribas}, I. and {Rieder}, M. and {Rohlfs}, R. and {Safa}, F. and {Salatti}, M. and {Santos}, N.~C. and {Scandariato}, G. and {S{\'e}gransan}, D. and {Simon}, A.~E. and {Smith}, A.~M.~S. and {Sordet}, M. and {Sousa}, S.~G. and {Steller}, M. and {Szab{\'o}}, G.~M. and {Szoke}, J. and {Thomas}, N. and {Tschentscher}, M. and {Udry}, S. and {Van Grootel}, V. and {Viotto}, V. and {Walter}, I. and {Walton}, N.~A. and {Wildi}, F. and {Wolter}, D.},
        title = "{The CHEOPS mission}",
      journal = {Experimental Astronomy},
         year = 2021,
        month = feb,
       volume = {51},
       number = {1},
        pages = {109-151},
          doi = {10.1007/s10686-020-09679-4},
archivePrefix = {arXiv},
       eprint = {2009.11633},
 primaryClass = {astro-ph.IM},
       adsurl = {https://ui.adsabs.harvard.edu/abs/2021ExA....51..109B}
}

@ARTICLE{Rauer2014,
       author = {{Rauer}, H. and {Catala}, C. and {Aerts}, C. and {Appourchaux}, T. and {Benz}, W. and {Brandeker}, A. and {Christensen-Dalsgaard}, J. and {Deleuil}, M. and {Gizon}, L. and {Goupil}, M. -J. and {G{\"u}del}, M. and {Janot-Pacheco}, E. and {Mas-Hesse}, M. and {Pagano}, I. and {Piotto}, G. and {Pollacco}, D. and {Santos}, {\.{C}}. and {Smith}, A. and {Su{\'a}rez}, J. -C. and {Szab{\'o}}, R. and {Udry}, S. and {Adibekyan}, V. and {Alibert}, Y. and {Almenara}, J. -M. and {Amaro-Seoane}, P. and {Eiff}, M. Ammler-von and {Asplund}, M. and {Antonello}, E. and {Barnes}, S. and {Baudin}, F. and {Belkacem}, K. and {Bergemann}, M. and {Bihain}, G. and {Birch}, A.~C. and {Bonfils}, X. and {Boisse}, I. and {Bonomo}, A.~S. and {Borsa}, F. and {Brand{\~a}o}, I.~M. and {Brocato}, E. and {Brun}, S. and {Burleigh}, M. and {Burston}, R. and {Cabrera}, J. and {Cassisi}, S. and {Chaplin}, W. and {Charpinet}, S. and {Chiappini}, C. and {Church}, R.~P. and {Csizmadia}, Sz. and {Cunha}, M. and {Damasso}, M. and {Davies}, M.~B. and {Deeg}, H.~J. and {D{\'\i}az}, R.~F. and {Dreizler}, S. and {Dreyer}, C. and {Eggenberger}, P. and {Ehrenreich}, D. and {Eigm{\"u}ller}, P. and {Erikson}, A. and {Farmer}, R. and {Feltzing}, S. and {de Oliveira Fialho}, F. and {Figueira}, P. and {Forveille}, T. and {Fridlund}, M. and {Garc{\'\i}a}, R.~A. and {Giommi}, P. and {Giuffrida}, G. and {Godolt}, M. and {Gomes da Silva}, J. and {Granzer}, T. and {Grenfell}, J.~L. and {Grotsch-Noels}, A. and {G{\"u}nther}, E. and {Haswell}, C.~A. and {Hatzes}, A.~P. and {H{\'e}brard}, G. and {Hekker}, S. and {Helled}, R. and {Heng}, K. and {Jenkins}, J.~M. and {Johansen}, A. and {Khodachenko}, M.~L. and {Kislyakova}, K.~G. and {Kley}, W. and {Kolb}, U. and {Krivova}, N. and {Kupka}, F. and {Lammer}, H. and {Lanza}, A.~F. and {Lebreton}, Y. and {Magrin}, D. and {Marcos-Arenal}, P. and {Marrese}, P.~M. and {Marques}, J.~P. and {Martins}, J. and {Mathis}, S. and {Mathur}, S. and {Messina}, S. and {Miglio}, A. and {Montalban}, J. and {Montalto}, M. and {Monteiro}, M.~J.~P.~F.~G. and {Moradi}, H. and {Moravveji}, E. and {Mordasini}, C. and {Morel}, T. and {Mortier}, A. and {Nascimbeni}, V. and {Nelson}, R.~P. and {Nielsen}, M.~B. and {Noack}, L. and {Norton}, A.~J. and {Ofir}, A. and {Oshagh}, M. and {Ouazzani}, R. -M. and {P{\'a}pics}, P. and {Parro}, V.~C. and {Petit}, P. and {Plez}, B. and {Poretti}, E. and {Quirrenbach}, A. and {Ragazzoni}, R. and {Raimondo}, G. and {Rainer}, M. and {Reese}, D.~R. and {Redmer}, R. and {Reffert}, S. and {Rojas-Ayala}, B. and {Roxburgh}, I.~W. and {Salmon}, S. and {Santerne}, A. and {Schneider}, J. and {Schou}, J. and {Schuh}, S. and {Schunker}, H. and {Silva-Valio}, A. and {Silvotti}, R. and {Skillen}, I. and {Snellen}, I. and {Sohl}, F. and {Sousa}, S.~G. and {Sozzetti}, A. and {Stello}, D. and {Strassmeier}, K.~G. and {{\v{S}}vanda}, M. and {Szab{\'o}}, Gy. M. and {Tkachenko}, A. and {Valencia}, D. and {Van Grootel}, V. and {Vauclair}, S.~D. and {Ventura}, P. and {Wagner}, F.~W. and {Walton}, N.~A. and {Weingrill}, J. and {Werner}, S.~C. and {Wheatley}, P.~J. and {Zwintz}, K.},
        title = "{The PLATO 2.0 mission}",
      journal = {Experimental Astronomy},
         year = 2014,
        month = nov,
       volume = {38},
       number = {1-2},
        pages = {249-330},
          doi = {10.1007/s10686-014-9383-4},
archivePrefix = {arXiv},
       eprint = {1310.0696},
 primaryClass = {astro-ph.EP},
       adsurl = {https://ui.adsabs.harvard.edu/abs/2014ExA....38..249R}
}

@ARTICLE{Ge2022,
       author = {{Ge}, Jian and {Zhang}, Hui and {Zang}, Weicheng and {Deng}, Hongping and {Mao}, Shude and {Xie}, Ji-Wei and {Liu}, Hui-Gen and {Zhou}, Ji-Lin and {Willis}, Kevin and {Huang}, Chelsea and {Howell}, Steve B. and {Feng}, Fabo and {Zhu}, Jiapeng and {Yao}, Xinyu and {Liu}, Beibei and {Aizawa}, Masataka and {Zhu}, Wei and {Li}, Ya-Ping and {Ma}, Bo and {Ye}, Quanzhi and {Yu}, Jie and {Xiang}, Maosheng and {Yu}, Cong and {Liu}, Shangfei and {Yang}, Ming and {Wang}, Mu-Tian and {Shi}, Xian and {Fang}, Tong and {Zong}, Weikai and {Liu}, Jinzhong and {Zhang}, Yu and {Zhang}, Liyun and {El-Badry}, Kareem and {Shen}, Rongfeng and {Tam}, Pak-Hin Thomas and {Hu}, Zhecheng and {Yang}, Yanlv and {Zou}, Yuan-Chuan and {Wu}, Jia-Li and {Lei}, Wei-Hua and {Wei}, Jun-Jie and {Wu}, Xue-Feng and {Sun}, Tian-Rui and {Wang}, Fa-Yin and {Zhang}, Bin-Bin and {Xu}, Dong and {Yang}, Yuan-Pei and {Li}, Wen-Xiong and {Xiang}, Dan-Feng and {Wang}, Xiaofeng and {Wang}, Tinggui and {Zhang}, Bing and {Jia}, Peng and {Yuan}, Haibo and {Zhang}, Jinghua and {Xuesong Wang}, Sharon and {Gan}, Tianjun and {Wang}, Wei and {Zhao}, Yinan and {Liu}, Yujuan and {Wei}, Chuanxin and {Kang}, Yanwu and {Yang}, Baoyu and {Qi}, Chao and {Liu}, Xiaohua and {Zhang}, Quan and {Zhu}, Yuji and {Zhou}, Dan and {Zhang}, Congcong and {Yu}, Yong and {Zhang}, Yongshuai and {Li}, Yan and {Tang}, Zhenghong and {Wang}, Chaoyan and {Wang}, Fengtao and {Li}, Wei and {Cheng}, Pengfei and {Shen}, Chao and {Li}, Baopeng and {Pan}, Yue and {Yang}, Sen and {Gao}, Wei and {Song}, Zongxi and {Wang}, Jian and {Zhang}, Hongfei and {Chen}, Cheng and {Wang}, Hui and {Zhang}, Jun and {Wang}, Zhiyue and {Zeng}, Feng and {Zheng}, Zhenhao and {Zhu}, Jie and {Guo}, Yingfan and {Zhang}, Yihao and {Li}, Yudong and {Wen}, Lin and {Feng}, Jie and {Chen}, Wen and {Chen}, Kun and {Han}, Xingbo and {Yang}, Yingquan and {Wang}, Haoyu and {Duan}, Xuliang and {Huang}, Jiangjiang and {Liang}, Hong and {Bi}, Shaolan and {Gai}, Ning and {Ge}, Zhishuai and {Guo}, Zhao and {Huang}, Yang and {Li}, Gang and {Li}, Haining and {Li}, Tanda and {Yuxi} and {Lu} and {Rix}, Hans-Walter and {Shi}, Jianrong and {Song}, Fen and {Tang}, Yanke and {Ting}, Yuan-Sen and {Wu}, Tao and {Wu}, Yaqian and {Yang}, Taozhi and {Yin}, Qing-Zhu and {Gould}, Andrew and {Lee}, Chung-Uk and {Dong}, Subo and {Yee}, Jennifer C. and {Shvartzvald}, Yossi and {Yang}, Hongjing and {Kuang}, Renkun and {Zhang}, Jiyuan and {Liao}, Shilong and {Qi}, Zhaoxiang and {Yang}, Jun and {Zhang}, Ruisheng and {Jiang}, Chen and {Ou}, Jian-Wen and {Li}, Yaguang and {Beck}, Paul and {Bedding}, Timothy R. and {Campante}, Tiago L. and {Chaplin}, William J. and {Christensen-Dalsgaard}, J{\o}rgen and {Garc{\'\i}a}, Rafael A. and {Gaulme}, Patrick and {Gizon}, Laurent and {Hekker}, Saskia and {Huber}, Daniel and {Khanna}, Shourya and {Li}, Yan and {Mathur}, Savita and {Miglio}, Andrea and {Mosser}, Beno{\^\i}t and {Ong}, J.~M. Joel and {Santos}, {\^A}ngela R.~G. and {Stello}, Dennis and {Bowman}, Dominic M. and {Lares-Martiz}, Mariel and {Murphy}, Simon and {Niu}, Jia-Shu and {Ma}, Xiao-Yu and {Moln{\'a}r}, L{\'a}szl{\'o} and {Fu}, Jian-Ning and {De Cat}, Peter and {Su}, Jie and {consortium}, the ET},
        title = "{ET White Paper: To Find the First Earth 2.0}",
      journal = {arXiv e-prints},
         year = 2022,
        month = jun,
          eid = {arXiv:2206.06693},
        pages = {arXiv:2206.06693},
          doi = {10.48550/arXiv.2206.06693},
archivePrefix = {arXiv},
       eprint = {2206.06693},
 primaryClass = {astro-ph.IM},
       adsurl = {https://ui.adsabs.harvard.edu/abs/2022arXiv220606693G}
}

@ARTICLE{TeacheyKipping2018,
       author = {{Teachey}, Alex and {Kipping}, David M.},
        title = "{Evidence for a large exomoon orbiting Kepler-1625b}",
      journal = {Science Advances},
         year = 2018,
        month = oct,
       volume = {4},
       number = {10},
          eid = {eaav1784},
        pages = {eaav1784},
          doi = {10.1126/sciadv.aav1784},
archivePrefix = {arXiv},
       eprint = {1810.02362},
 primaryClass = {astro-ph.EP},
       adsurl = {https://ui.adsabs.harvard.edu/abs/2018SciA....4.1784T}
}

@ARTICLE{Yahalomi2024,
       author = {{Yahalomi}, Daniel A. and {Kipping}, David and {Nesvorn{\'y}}, David and {Dalba}, Paul A. and {Benni}, Paul and {Cacho-Negrete}, Ceiligh and {Collins}, Karen and {Earwicker}, Joel T. and {Lewis}, John Arban and {McLeod}, Kim K. and {Schwarz}, Richard P. and {Wang}, Gavin},
        title = "{Not-so-fast Kepler-1513: a perturbing planetary interloper in the exomoon corridor}",
      journal = {\mnras},
         year = 2024,
        month = jan,
       volume = {527},
       number = {1},
        pages = {620-639},
          doi = {10.1093/mnras/stad3070},
archivePrefix = {arXiv},
       eprint = {2310.03802},
 primaryClass = {astro-ph.EP},
       adsurl = {https://ui.adsabs.harvard.edu/abs/2024MNRAS.527..620Y}
}

@book{ensemble_learning,
author = {Zhang, C. and Ma, Y.},
year = {2012},
month = {01},
pages = {1-329},
title = {Ensemble machine learning: Methods and applications},
doi = {10.1007/9781441993267}
}

@ARTICLE{DasarathySheela1979,
  author={Dasarathy, B.V. and Sheela, B.V.},
  journal={Proceedings of the IEEE}, 
  title={A composite classifier system design: Concepts and methodology}, 
  year={1979},
  volume={67},
  number={5},
  pages={708-713},
  doi={10.1109/PROC.1979.11321}}

@ARTICLE{HansenSalamon1990,
  author={Hansen, L.K. and Salamon, P.},
  journal={IEEE Transactions on Pattern Analysis and Machine Intelligence}, 
  title={Neural network ensembles}, 
  year={1990},
  volume={12},
  number={10},
  pages={993-1001},
  doi={10.1109/34.58871}}

@article{Schapire1990,
title = "The Strength of Weak Learnability",
author = "Schapire, {Robert E.}",
year = "1990",
month = jun,
doi = "10.1023/A:1022648800760",
language = "English (US)",
volume = "5",
pages = "197--227",
journal = "Machine Learning",
issn = "0885-6125",
publisher = "Springer Netherlands",
number = "2",
}

@ARTICLE{Kipping2013c,
       author = {{Kipping}, D.~M. and {Hartman}, J. and {Buchhave}, L.~A. and {Schmitt}, A.~R. and {Bakos}, G. {\'A}. and {Nesvorn{\'y}}, D.},
        title = "{The Hunt for Exomoons with Kepler (HEK). II. Analysis of Seven Viable Satellite-hosting Planet Candidates}",
      journal = {\apj},
         year = 2013,
        month = jun,
       volume = {770},
       number = {2},
          eid = {101},
        pages = {101},
          doi = {10.1088/0004-637X/770/2/101},
archivePrefix = {arXiv},
       eprint = {1301.1853},
 primaryClass = {astro-ph.EP},
       adsurl = {https://ui.adsabs.harvard.edu/abs/2013ApJ...770..101K}
}

@ARTICLE{Fabrycky2012,
       author = {{Fabrycky}, Daniel C. and {Ford}, Eric B. and {Steffen}, Jason H. and {Rowe}, Jason F. and {Carter}, Joshua A. and {Moorhead}, Althea V. and {Batalha}, Natalie M. and {Borucki}, William J. and {Bryson}, Steve and {Buchhave}, Lars A. and {Christiansen}, Jessie L. and {Ciardi}, David R. and {Cochran}, William D. and {Endl}, Michael and {Fanelli}, Michael N. and {Fischer}, Debra and {Fressin}, Francois and {Geary}, John and {Haas}, Michael R. and {Hall}, Jennifer R. and {Holman}, Matthew J. and {Jenkins}, Jon M. and {Koch}, David G. and {Latham}, David W. and {Li}, Jie and {Lissauer}, Jack J. and {Lucas}, Philip and {Marcy}, Geoffrey W. and {Mazeh}, Tsevi and {McCauliff}, Sean and {Quinn}, Samuel and {Ragozzine}, Darin and {Sasselov}, Dimitar and {Shporer}, Avi},
        title = "{Transit Timing Observations from Kepler. IV. Confirmation of Four Multiple-planet Systems by Simple Physical Models}",
      journal = {\apj},
         year = 2012,
        month = may,
       volume = {750},
       number = {2},
          eid = {114},
        pages = {114},
          doi = {10.1088/0004-637X/750/2/114},
archivePrefix = {arXiv},
       eprint = {1201.5415},
 primaryClass = {astro-ph.EP},
       adsurl = {https://ui.adsabs.harvard.edu/abs/2012ApJ...750..114F}
}

@ARTICLE{Gautier2012,
       author = {{Gautier}, Thomas N., III and {Charbonneau}, David and {Rowe}, Jason F. and {Marcy}, Geoffrey W. and {Isaacson}, Howard and {Torres}, Guillermo and {Fressin}, Francois and {Rogers}, Leslie A. and {D{\'e}sert}, Jean-Michel and {Buchhave}, Lars A. and {Latham}, David W. and {Quinn}, Samuel N. and {Ciardi}, David R. and {Fabrycky}, Daniel C. and {Ford}, Eric B. and {Gilliland}, Ronald L. and {Walkowicz}, Lucianne M. and {Bryson}, Stephen T. and {Cochran}, William D. and {Endl}, Michael and {Fischer}, Debra A. and {Howell}, Steve B. and {Horch}, Elliott P. and {Barclay}, Thomas and {Batalha}, Natalie and {Borucki}, William J. and {Christiansen}, Jessie L. and {Geary}, John C. and {Henze}, Christopher E. and {Holman}, Matthew J. and {Ibrahim}, Khadeejah and {Jenkins}, Jon M. and {Kinemuchi}, Karen and {Koch}, David G. and {Lissauer}, Jack J. and {Sanderfer}, Dwight T. and {Sasselov}, Dimitar D. and {Seager}, Sara and {Silverio}, Kathryn and {Smith}, Jeffrey C. and {Still}, Martin and {Stumpe}, Martin C. and {Tenenbaum}, Peter and {Van Cleve}, Jeffrey},
        title = "{Kepler-20: A Sun-like Star with Three Sub-Neptune Exoplanets and Two Earth-size Candidates}",
      journal = {\apj},
         year = 2012,
        month = apr,
       volume = {749},
       number = {1},
          eid = {15},
        pages = {15},
          doi = {10.1088/0004-637X/749/1/15},
archivePrefix = {arXiv},
       eprint = {1112.4514},
 primaryClass = {astro-ph.EP},
       adsurl = {https://ui.adsabs.harvard.edu/abs/2012ApJ...749...15G}
}

@ARTICLE{Giles2018,
       author = {{Giles}, H.~A.~C. and {Bayliss}, D. and {Espinoza}, N. and {Brahm}, R. and {Blanco-Cuaresma}, S. and {Shporer}, A. and {Armstrong}, D. and {Lovis}, C. and {Udry}, S. and {Bouchy}, F. and {Marmier}, M. and {Jord{\'a}n}, A. and {Bento}, J. and {Collier Cameron}, A. and {Sefako}, R. and {Cochran}, W.~D. and {Rojas}, F. and {Rabus}, M. and {Jenkins}, J.~S. and {Jones}, M. and {Pantoja}, B. and {Soto}, M. and {Jensen-Clem}, R. and {Duev}, D.~A. and {Salama}, M. and {Riddle}, R. and {Baranec}, C. and {Law}, N.~M.},
        title = "{K2-140b - an eccentric 6.57 d transiting hot Jupiter in Virgo}",
      journal = {\mnras},
         year = 2018,
        month = apr,
       volume = {475},
       number = {2},
        pages = {1809-1818},
          doi = {10.1093/mnras/stx3300},
archivePrefix = {arXiv},
       eprint = {1706.06865},
 primaryClass = {astro-ph.EP},
       adsurl = {https://ui.adsabs.harvard.edu/abs/2018MNRAS.475.1809G}
}

@article{DurbinWatson1950,
    author = {Durbin, J. and Watson, G. S.},
    title = "{TESTING FOR SERIAL CORRELATION IN LEAST SQUARES REGRESSION. I}",
    journal = {Biometrika},
    volume = {37},
    number = {3-4},
    pages = {409-428},
    year = {1950},
    month = {12},
    issn = {0006-3444},
    doi = {10.1093/biomet/37.3-4.409},
    url = {https://doi.org/10.1093/biomet/37.3-4.409},
    eprint = {https://academic.oup.com/biomet/article-pdf/37/3-4/409/422190/37-3-4-409.pdf},
}

@article{celerite1,
   author = {{Foreman-Mackey}, D. and {Agol}, E. and {Ambikasaran}, S. and
            {Angus}, R.},
    title = "{Fast and Scalable Gaussian Process Modeling with Applications to
              Astronomical Time Series}",
  journal = {\aj},
     year = 2017,
    month = dec,
   volume = 154,
    pages = {220},
      doi = {10.3847/1538-3881/aa9332},
   adsurl = {http://adsabs.harvard.edu/abs/2017AJ....154..220F}
}

@article{celerite2,
   author = {{Foreman-Mackey}, D.},
    title = "{Scalable Backpropagation for Gaussian Processes using Celerite}",
  journal = {Research Notes of the American Astronomical Society},
     year = 2018,
    month = feb,
   volume = 2,
   number = 1,
    pages = {31},
      doi = {10.3847/2515-5172/aaaf6c},
   adsurl = {http://adsabs.harvard.edu/abs/2018RNAAS...2a..31F}
}

@ARTICLE{Sandford2017,
       author = {{Sandford}, Emily and {Kipping}, David},
        title = "{Know the Planet, Know the Star: Precise Stellar Densities from Kepler Transit Light Curves}",
      journal = {\aj},
         year = 2017,
        month = dec,
       volume = {154},
       number = {6},
          eid = {228},
        pages = {228},
          doi = {10.3847/1538-3881/aa94bf},
archivePrefix = {arXiv},
       eprint = {1710.07293},
 primaryClass = {astro-ph.EP},
       adsurl = {https://ui.adsabs.harvard.edu/abs/2017AJ....154..228S}
}

@article{Schwarz1978,
 ISSN = {00905364},
 URL = {http://www.jstor.org/stable/2958889},
 author = {Gideon Schwarz},
 journal = {The Annals of Statistics},
 number = {2},
 pages = {461--464},
 publisher = {Institute of Mathematical Statistics},
 title = {Estimating the Dimension of a Model},
 urldate = {2022-09-16},
 volume = {6},
 year = {1978}
}

@ARTICLE{Galton1907,
       author = {{Galton}, Francis},
        title = "{Vox Populi}",
      journal = {Nature},
         year = 1907,
        month = mar,
       volume = {75},
       number = {1949},
        pages = {450-451},
          doi = {10.1038/075450a0},
       adsurl = {https://ui.adsabs.harvard.edu/abs/1907Natur..75..450G}
}
\bibliographystyle{aasjournal}




\end{document}